\documentclass[sigconf,nonacm]{acmart}

\renewcommand\footnotetextcopyrightpermission[1]{}
\usepackage{amsmath}          
\usepackage{graphicx}
\usepackage{grffile}
\usepackage{booktabs,multirow}
\usepackage{subcaption}
\usepackage{algorithm}
\usepackage{algpseudocode}
\usepackage{comment}
\usepackage{tikz}
\usepackage{hyperref}
\hypersetup{colorlinks,allcolors=black}

\usetikzlibrary{arrows.meta,calc,positioning}
\newcommand*\circled[1]{\tikz[baseline=(char.base)]{
    \node[shape=circle,draw,inner sep=1pt] (char) {#1};}}
    
\title{Hybrid SLC-MLC RRAM Mixed-Signal Processing-in-Memory Architecture for Transformer Acceleration via Gradient Redistribution}

\author{Chang Eun Song}
\affiliation{%
  \institution{University of California, San Diego}
  \city{San Diego}\state{CA}\country{USA}}
\email{cesong@ucsd.edu}

\author{Priyansh Bhatnagar}
\affiliation{%
  \institution{University of California, San Diego}
  \city{San Diego}\state{CA}\country{USA}}
\email{prbhatnagar@ucsd.edu}

\author{Zihan Xia}
\affiliation{%
  \institution{University of California, San Diego}
  \city{San Diego}\state{CA}\country{USA}}
\email{z5xia@ucsd.edu}

\author{Nam Sung Kim}
\affiliation{%
  \institution{University of Illinois at Urbana–Champaign}
  \city{Urbana}\state{IL}\country{USA}}
\email{nam.sung.kim@gmail.com}

\author{Tajana S.~Rosing}
\affiliation{%
  \institution{University of California, San Diego}
  \city{San Diego}\state{CA}\country{USA}}
\email{tajana@ucsd.edu}

\author{Mingu Kang}
\affiliation{%
  \institution{University of California, San Diego}
  \city{San Diego}\state{CA}\country{USA}}
\email{mingu@ucsd.edu}

\begin{document}

\begin{abstract}
Transformers, while revolutionary, face challenges due to their demanding computational cost and large data movement. To address this, we propose HyFlexPIM, a novel  mixed-signal processing-in-memory (PIM) accelerator for inference that flexibly utilizes both single-level cell (SLC) and multi-level cell (MLC) RRAM technologies to trade-off accuracy and efficiency. HyFlexPIM achieves efficient dual-mode operation by utilizing digital PIM for high-precision and write-intensive operations while analog PIM for high parallel and low-precision computations. The analog PIM further distributes tasks between SLC and MLC PIM operations, where a single analog PIM module can be reconfigured to switch between two operations (SLC/MLC) with minimal overhead (<1\% for area \& energy). Critical weights are allocated to SLC RRAM for high accuracy, while less critical  weights are assigned to MLC RRAM to maximize capacity, power, and latency efficiency. However, despite employing such a hybrid mechanism, brute-force mapping on hardware fails to deliver significant benefits due to the limited proportion of weights accelerated by the MLC and the noticeable degradation in accuracy.
To maximize the potential of our hybrid hardware architecture, we propose an algorithm co-optimization technique, called \textit{gradient redistribution}, which uses Singular Value Decomposition (SVD) to decompose and truncate matrices based on their importance, then fine-tune them to concentrate significance into a small subset of weights. By doing so, only 5-10\% of the weights have dominantly large gradients, making it favorable for HyFlexPIM by minimizing the use of expensive SLC RRAM while maximizing the efficient MLC RRAM. Our evaluation shows that HyFlexPIM significantly enhances computational throughput and energy efficiency, achieving maximum 1.86$\times$ and 1.45$\times$ higher than state-of-the-art methods.
\end{abstract}


\maketitle    

\section{Introduction}

Transformer models have brought significant advancements in a wide range of machine learning (ML) applications, including natural language processing and computer vision~\cite{topal2021exploring}.  However, their high computational cost and memory bandwidth/capacity requirements pose considerable challenges for the hardware deployment, given the limited resources such as computing capability, memory bandwidth, energy, and delay budgets~\cite{lu2020hardware, wang2020hat}. These difficulties arise from the immense memory bandwidth usage and intensive data transfer cost required for model operations, making it challenging to process them efficiently with conventional accelerators such as GPUs and ASICs \cite{10.1145/3579371.3589057, yazdanbakhsh2022sparse, 10.1145/3665314.3670798, yang2024fsl, song202452}.
As an alternative solution, memory-centric processing architectures have been emerging, specifically processing-in-memory (PIM) \cite{zidan2018future, yao2020fully, ren2023associative}. These technologies eliminate data movement and leverage memory's inherent parallelism to enhance computational efficiency. In particular, both digital and analog PIM approaches have shown significant efficiency improvements - digital PIM reduces data movement costs while maintaining the accuracy for high bit-precision operations \cite{jin2021rehy, talati2016logic, truong2021racer, gao2019computedram, zhou2022transpim, wu2024pim, liu2023hardsea}, and analog PIM demonstrates dramatic efficiency improvement, achieving more than two orders of magnitude benefit \cite{JSSC_SGD, jin2021rehy, andrulis2023raella, cao2021neural} through low-voltage swing operations.


To leverage such opportunities, several efforts have explored PIM techniques that employ  DRAM, SRAM,  RRAM (Resistive RAM), PRAM (Phase Change Memory), and other emerging memory topologies \cite{ali2019memory, smullen2011relaxing, yang2023processing}. Out of these efforts, RRAM is one of the most widely explored devices for PIM acceleration due to its non-volatile memory (NVM) nature, high storage density, fast read operation, low energy consumption, and excellent analog programmability for the multi-level cell (MLC) \cite{zhang2020neuro, zidan2018future, yao2020fully, prezioso2015training, jiang2019circuit, ramadan2019adaptive, talati2016logic, truong2021racer}. Despite its great potential in terms of efficiency, practical applications of RRAM for Transformer models face multiple challenges in terms of its non-ideal characteristics and reliability concerns. While digital RRAM PIM approaches have been actively explored for various complex models due to their reliability and precision advantages \cite{talati2016logic, truong2021racer, gao2019computedram}, analog RRAM PIM has been explored mostly in simpler algorithms such as multilayer perceptrons, spiking neural networks (SNNs), and convolutional neural networks (CNNs), which exhibit relatively high inherent error resiliency within the model itself \cite{liu2023area, kim2023samba, xue202116}.


On the other hand, Transformer models consist of many cascaded layers and often require high bit precision for specific operations, making them sensitive to noise sources and resulting in immediate accuracy degradation from such noise. Due to these accuracy concerns, a large body of work \cite{zhou2022transpim, wu2024pim, liu2023hardsea} has explored digital PIM solutions for processing Transformers while reducing data movement costs. However, it has shown limited improvement due to less parallelism in General matrix-vector multiplications (GEMV) operations than analog PIMs \cite{yang2020retransformer}, falling short of  fully leveraging the substantial potential for energy and delay efficiency offered by low-swing analog operations.
SPIRNT \cite{yazdanbakhsh2022sparse} explored analog PIM within a limited scope, using it as a pre-processor before the main digital processing to filter redundant tokens by calculating the correlation scores between all query and key pairs in the RRAM array.
Despite a few analog efforts, the study of analog PIM acceleration for Transformers remains largely unexplored, mainly due to concerns about accuracy degradation.


One of the main advantages of analog RRAM PIM is its ability to support multi-level cell (MLC) programming, where multiple bits of information (2 to 6-bits) can be stored per cell based on programmable resistance levels, unlike single-level cell (SLC) which stores only a single bit (e.g., high vs. low). In addition, PIM with MLC achieves higher computing throughput as each memory cell processes multiple bits simultaneously. Despite above advantages, the inherent non-linearity in MLC operations leads to severe accuracy degradation. Our experiments demonstrate that even the simplest Transformer models \cite{devlin2018bert} suffer catastrophic accuracy loss when implemented entirely with 2-bit MLC, e.g., 40\% drop for BERT-Base with GLUE MRPC dataset.

In contrast, SLC offers a much higher noise margin against data distortion, ensuring better accuracy but at the cost of significantly greater area and energy overhead. Exploiting both types of devices (SLC \& MLC) in a single hardware offers a great knob to trade-off between efficiency and accuracy by containing the critical contents in the SLC while less important and error-tolerant data in the MLC. 
To unlock the true potential of enhanced efficiency and large capacity from analog PIM, it is desirable to process a dominant portion of data in the MLC.
However, applying these methods to Transformer encounters two major challenges: 1) 
the error-tolerant portion to be processed in MLC is often not sufficient enough so that the acceleration offered by the hybrid architecture does not lead to a significant improvement in efficiency, and 2)
even when this is not the case, it is hard to clearly demarcate which part is error-tolerant vs. error-susceptible. 


This work aims to tackle these challenges at the intersection of hardware \& software by introducing a novel hybrid SLC-MLC RRAM mixed-signal architecture, termed \textbf{HyFlexPIM} along with an algorithmic transformation called \textit{gradient redistribution}. HyFlexPIM exhibits three key characteristics (1) \textbf{mixed-signal}, leveraging both digital \& analog \textbf{\underline{PIM}} computations, (2) \textbf{\underline{Hy}brid}, further dividing the analog PIM into SLC and MLC RRAM operations given the accuracy requirement, and (3) \textbf{\underline{Flex}ible}, enabling seamless switching between SLC-MLC configurations in a single PIM unit.

On the software side, our proposed gradient redistribution technique reshapes Transformer models without accuracy compromise to be more compatible with the hybrid hardware platform, rather than passively relying on the inherent error resiliency of the model itself. In this approach, we accomplish these goals by employing Singular Value Decomposition (SVD) to decompose matrices based on their order of importance. After truncating the post-SVD matrix to remove redundancy, we fine-tune it so that a small portion of the weights have dominant importance. 
This  approach will not only facilitate a clear demarcation between critical and error-tolerant  sections of the task but will also enhance the size of the error-tolerant section for efficient hardware mapping, thereby maximizing the use of low-power analog processing in the MLC while achieving higher storage density and improved computational throughput. 
 All SVD decomposition, truncation, and fine-tuning are performed entirely in the software in advance. Meanwhile, the hardware, featuring reconfigurability for hybrid SLC-MLC operations, only utilizes the final truncated matrices for GEMV operations for inference, resulting in lower computing and storage costs.

Our detailed contributions are as follows:

\noindent $\bullet$ We introduce architectural and circuit-level hardware designs to support the dual-tier approach, facilitating importance-based data flow with hybrid SLC-MLC RRAM mixed-signal PIM and a reconfigurable ADC for dynamic precision.

\noindent $\bullet$ The gradient redistribution technique allows us to protect only the most important 5-10\% \& 5-20\% of weights by storing them in the SLC for the encoder \& decoder models, respectively, to achieve high precision while most weights are processed in the MLC.

\noindent $\bullet$ We performed accuracy analysis by employing realistic circuit error models to reflect the non-idealities of real RRAM devices \cite{fan2024efficientopenmodificationspectral}, peripheral circuitry, and the detailed computational and data flow.

\noindent $\bullet$ We conducted extensive experiments on various benchmarks, including both encoder-based, decoder-based, and vision transformer models such as BERT \cite{devlin2018bert}, GPT-2 \cite{radford2019language}, Llama3 \cite{llama3.2}, and ViT-Base \cite{dosovitskiy2021image} with multiple datasets of GLUE (7 tasks) \cite{wang2019glue}, WikiText-2 \cite{wikitext2}, PTB \cite{marcus-etal-1993-building}, and CIFAR-10 \cite{Krizhevsky09learningmultiple}. 

The results indicate that HyFlexPIM exhibits superior throughput and energy efficiency of maximum 1.86$\times$ and 1.24$\times$ compared to state-of-the-art (SoA) methods \cite{li2024asadi}, with negligible accuracy repercussions (less than 1\%) in the encoder model and vision transformer and loss increase (less than 10\% of loss) in decoder model, on average.



\section{Background and Motivation}

 \subsection{Transformer} \label{subsec:Transformer}

\begin{figure}[t]
\centering
\includegraphics[width=0.8\columnwidth]{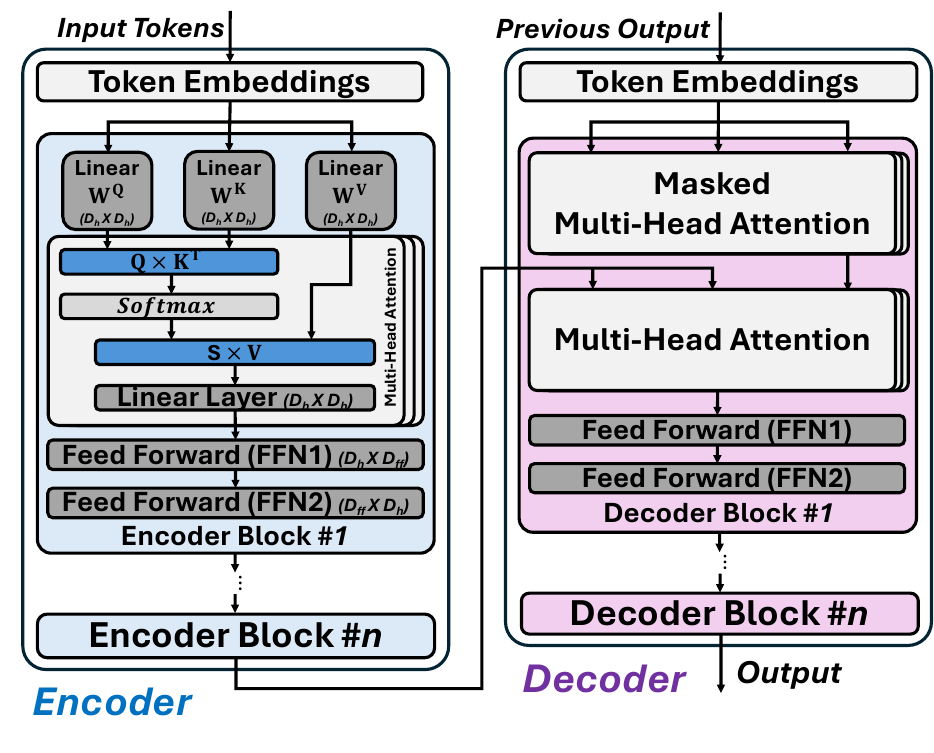} 
\caption{Variants of Transformer model architectures and dataflow: (a)  an encoder and (b) a decoder architecture.}
\label{fig:transformer}
\end{figure} 

Figure \ref{fig:transformer} illustrates the computing stages of an end-to-end Transformer encoder and decoder, which consist of multiple blocks internally. 
Attention-based models fundamentally consist of two primary modules: Multi-Head Attention (MHA) and Feed Forward Network (FFN). The attention mechanism is designed to selectively focus on a subset of segments in the input sequence while generating an output. The core concept involves computing a weighted sum of  values (\(\mathbf{V}\)), where the weights are determined by the similarity between a query (\(\mathbf{Q}\)) and a set of keys (\(\mathbf{K}\)). This process can be mathematically represented as follows: 
\begin{equation}
\mathbf{Q} = \mathbf{W^Q}(x), \quad \mathbf{K} = \mathbf{W^K}(x), \quad \mathbf{V} = \mathbf{W^V}(x)
\end{equation}
where \(\mathbf{Q}\), \(\mathbf{K}\), and \(\mathbf{V}\) are \(L \times D_h\) dimensional matrices with \(L\) representing the sequence length, \(D_h\) as hidden dimension, and \(x\) as input embedding.
Here, \(\mathbf{W^Q}\), \(\mathbf{W^K}\), and \(\mathbf{W^V}\) are learnable weight matrices for the generation of queries, keys, and values, respectively.
Then, the attention scores are computed as the dot product between \(\mathbf{Q}\) and \(\mathbf{K}\), followed by the application of a softmax function to obtain the normalized weights. The attention mechanism is formulated as follows with \(d_{head}\) being  the size of each head:
\begin{equation}
\text{Attn}(\mathbf{Q}, \mathbf{K}, \mathbf{V}) = \text{softmax}\left(\frac{\mathbf{Q}\mathbf{K}^T}{\sqrt{d_{head}}}\right)\mathbf{V}
\label{eq:qkv}
\end{equation}
Typically, the attention module is followed by an additional linear layer with a weight matrix $\mathbf{W_{proj}}$ for further processing.
After MHA, the output is processed through a FFN consisting of two layers: FFN1 and FFN2. FFN1 performs a linear transformation using a weight matrix $\mathbf{W_{ffn1}}$ of dimensions \(D_h \times D_{ff}\), followed by a non-linear activation function, typically ReLU or GELU. This layer projects the \(D_h\)-dimensional input to a higher-dimensional space of size \(D_{ff}\), resulting in an output dimension of \(L \times D_{ff}\).
Subsequently, FFN2 applies a linear transformation to project the intermediate representation back to the original hidden dimension size of  \(L \times D_h\) with a weight matrix $\mathbf{W_{ffn2}}$ with  dimension of \(D_{ff} \times D_h\).  

\begin{figure}[t]
\centering
\includegraphics[width=0.9\columnwidth]{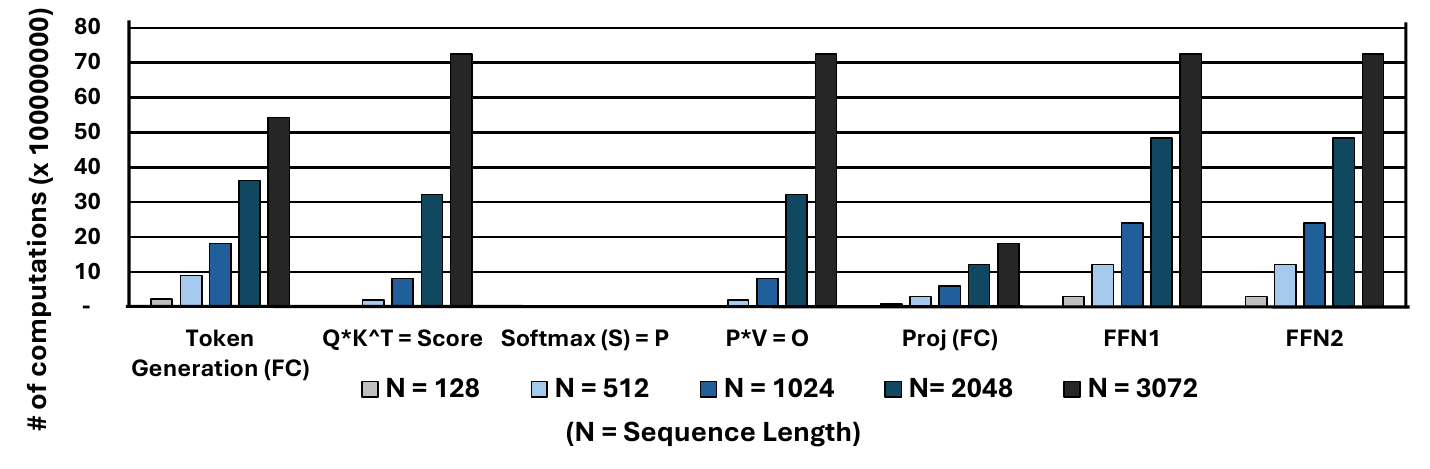}
\caption{The number of operations for each stage in transformer models.}
\label{fig:comparam}
\end{figure}

Figure \ref{fig:comparam} shows the distribution of the number of operations for each computation stage in transformers  with respect to different sequence lengths. As a large portion (>70\%) of computations in Transformers is from static weights \cite{10.1145/3579371.3589057, 10447756} and does not update them during inference, the weights written in the array can be reused over a large number of inputs without writing, which is  costly due to the high writing voltage of NVM devices. For this reason,  most of the Transformer's computing pipeline can be efficiently accelerated by PIM.  Furthermore, this static nature enables pre-processing using singular value decomposition (SVD) (described in Section \ref{SVD_background}) for improved efficiency and supports the pre-storage of weights processed through SVD.

\subsection{Analog RRAM Processing In-Memory}

\begin{figure}[t]
\centering
\includegraphics[width=0.9\columnwidth]{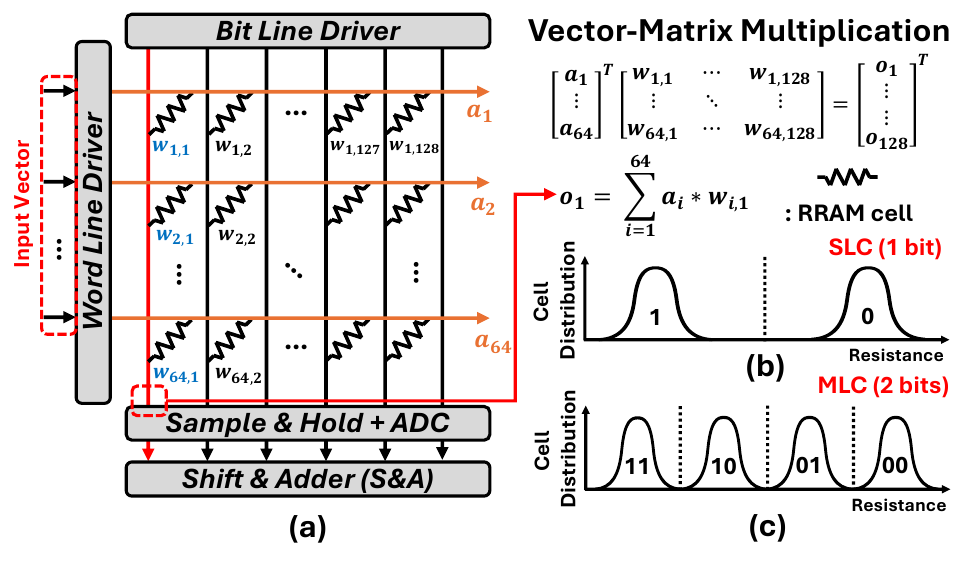}
\caption{Illustration of (a) analog RRAM processing-in-memory vector-matrix multiplication, (b) single-level cell (SLC), (c) multi-level cell (MLC) distribution, and (d) digital RRAM processing-in-memory NOR operation.}
\label{fig:RRAM}
\end{figure} 

Analog processing-in-memory (PIM) has been extensively adopted for accelerating a wide range of applications with high demands for memory and computational resources, exploiting its significant parallelism, low-power computation with low-voltage swing, and capability to minimize explicit data movement. While various memory topologies, including RRAM \cite{rram_isaac, rram_prime, yu2021rram, li2024asadi, yazdanbakhsh2022sparse, huang2023nonvolatile, liu2023hardsea}, PCM \cite{wong2010phase}, SRAM \cite{agrawal2019xcel}, and DRAM \cite{ali2019memory}, have been explored, this paper focuses on RRAM due to its suitability for both storage and computation. RRAM's non-volatile memory (NVM) nature, high storage density, low read power, and excellent analog programmability make it a highly suitable choice. Consequently, RRAM has been one of the most widely employed devices for the acceleration of highly parallel analog General matrix-vector multiplications (GEMV).

During GEMV PIM operations, each element of the matrix is mapped to the conductance ($=$1$/$resistance) of the memristor in a bitcell. While single-level cells (SLC) contain one bit per cell (Figure \ref{fig:RRAM}(b)), MLCs store multiple bits (e.g., 2 bits or more) per cell as multi-level conductance (Figure \ref{fig:RRAM}(c)).
Each element of input vector is applied to the all wordlines (WLs) of the RRAM array simultaneously, horizontal lines in Figure \ref{fig:RRAM}(a), by the WL driver as a voltage signal. 
When each row of the array receives one bit of the corresponding input element in a bit-sequential manner, and assuming SLC (each storing 1 bit), each bitcell performs 1-bit by 1-bit binary multiplication.
Conversely, if the cells are configured as MLCs, each bitcell performs 1-bit by multi-bit multiplications achieving higher throughput. The multiplication results from every cell are directly accumulated on the bitline (vertical line) of the array in the analog domain, generating the output of the GEMV.  The following equation formally represents the GEMV operation with an input vector $\mathbf{A}_{1 \times n}$ and matrix $\mathbf{W}_{n \times m}$ on an RRAM crossbar array:

\begin{equation}
o_{j} = \sum_{i=1}^n a_{i} \cdot w_{ij}
\end{equation}
where  
$w_{ij}$ is a conductance corresponding to the element at the $i$-th row and $j$-th column of the weight matrix $\mathbf{W}$ whereas $o_{j}$ is an $j$-th elements of output vector when an input vector (\{$a_{1}, ..., a_{n}$\}) was applied.
Finally, the accumulated analog current ($o_j$) is captured by a sample-and-hold circuit, and subsequently converted to digital form through an analog-to-digital converter (ADC). The final result of the multi-bit GEMV operation is obtained via weighted-sum of all output bits using a shift and add (S\&A) module to represent different bit positions of input. (Figure \ref{fig:RRAM} (a))


Despite its great potential for an efficient GEMV operation via full parallelism from all bitcells, practical applications of RRAM for Transformer models face multiple challenges due to its non-ideal characteristics and reliability concerns. These challenges include spatial mismatch from process variations, non-linear and asymmetric weight updates, limited endurance, and restricted data retention capabilities. Intuitively, SLC offers a higher noise margin against various noise sources \cite{fan2024efficientopenmodificationspectral} leading to better accuracy, but at the cost of significantly greater area, delay, and  energy overhead compared to MLC.

\subsection{Singular Value Decomposition (SVD)} \label{SVD_background}

\begin{figure}[t]
\centering
\includegraphics[width=0.9\columnwidth]{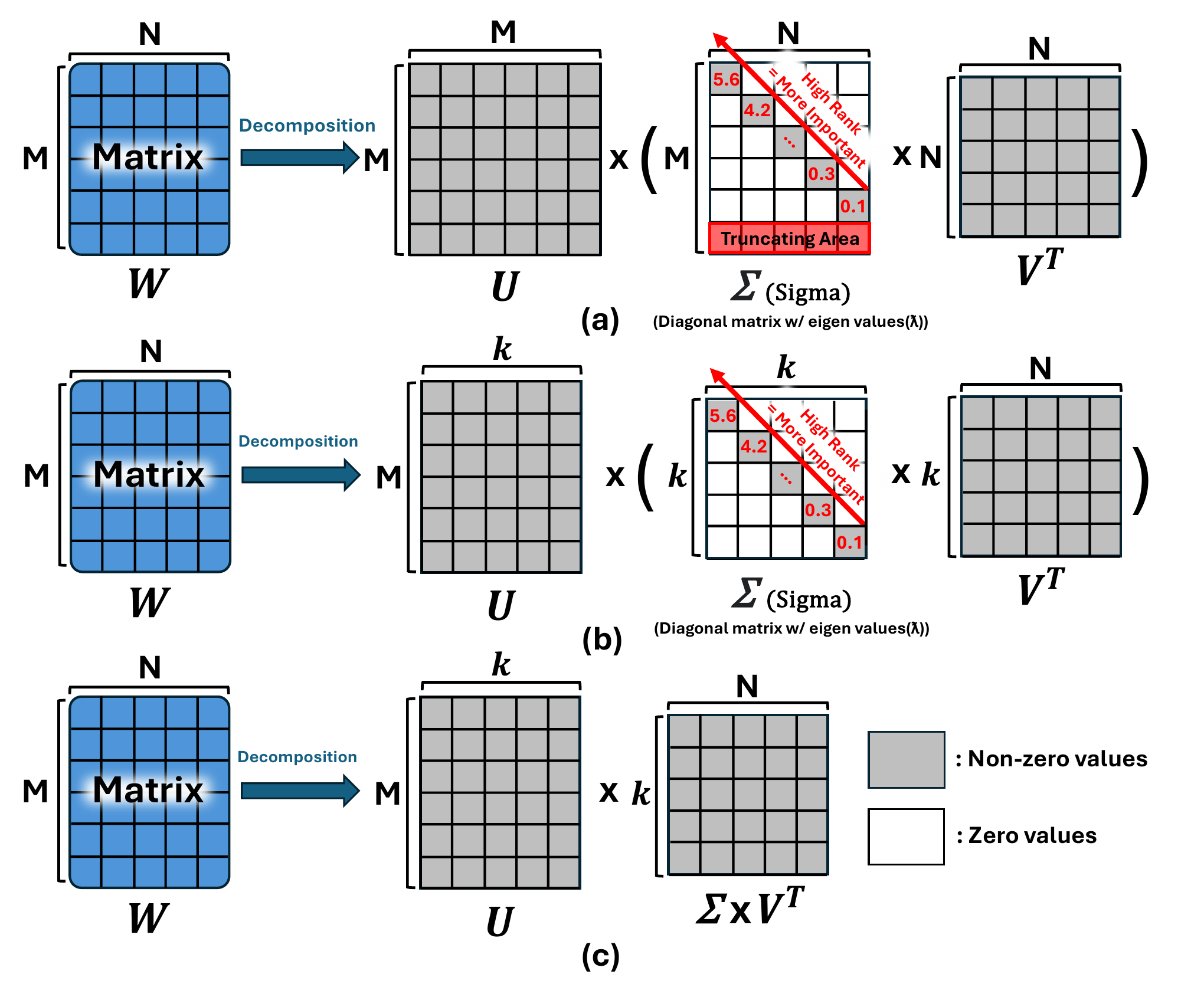}
\caption{Post-SVD matrix applied to static weights in the transformer (a) during training, (b) after truncation, and (c) 
during inference.}
\label{fig:svd}
\end{figure} 

SVD is a widely utilized method~\cite{abdi2007singular,lv2023lightformer} for decomposing weight matrices into three smaller matrices while retaining much of the original information. 
Any matrix $W \in \mathbb{R}^{M \times N}$ can be transformed as follows:
\begin{equation}
W = U \Sigma V^T
\end{equation}
As described in Figure \ref{fig:svd} (a), $ U $ and $ V^T $ are \( M \times M \) and \( N \times N \) orthogonal matrices, respectively, and \( \Sigma \) is an \( M \times N \) diagonal matrix with non-negative real numbers on the diagonal. This decomposition facilitates the analysis and manipulation of the matrix \( W \) based on the diagonal elements, known as singular values, of \( \Sigma \), which indicate the importance of the corresponding dimensions, e.g., the rows of $ U $ and the columns of $ V^T $.
To reduce dimensionality, as shown in Figure \ref{fig:svd} (b), a truncated SVD is commonly used: $W \approx W_k = U_k \Sigma_k V_k^T$, where $U_k$, $\Sigma_k$, and $V_k$ contain only the first $k$ columns, singular values, and rows, respectively. After truncation, $\Sigma_k V_k^T$ is pre-computed and stored as a matrix of size ($k \times N$), reducing both data volume and computation cost, as illustrated in Figure~\ref{fig:svd} (c). While these SVD processes are typically performed in software, the hardware focuses on storing and utilizing the truncated matrices for inference. Although SVD is widely employed for data compression through truncation, proposed algorithm repurpose it to reshape the matrices, maximizing the efficiency of the proposed hardware, as detailed in Section~\ref{SVD}.

\section{H\lowercase{y}F\lowercase{lex}PIM}

In this section, we introduce the detailed architecture of HyFlexPIM, which is a mixed-signal RRAM PIM accelerator designed to efficiently support both SLC and MLC PIM operations with negligible overhead.
This hybrid hardware creates a great synergy with our proposed \textit{gradient redistribution} technique to be detailed in the following Section~\ref{SVD}, enabling effective mapping of critical weights to SLC and less important weights to MLC.

\subsection{HyFlexPIM Hardware Architecture}\label{sec: hardware_arch}

\begin{figure*}[th]
\centering
\includegraphics[width=1.8\columnwidth]{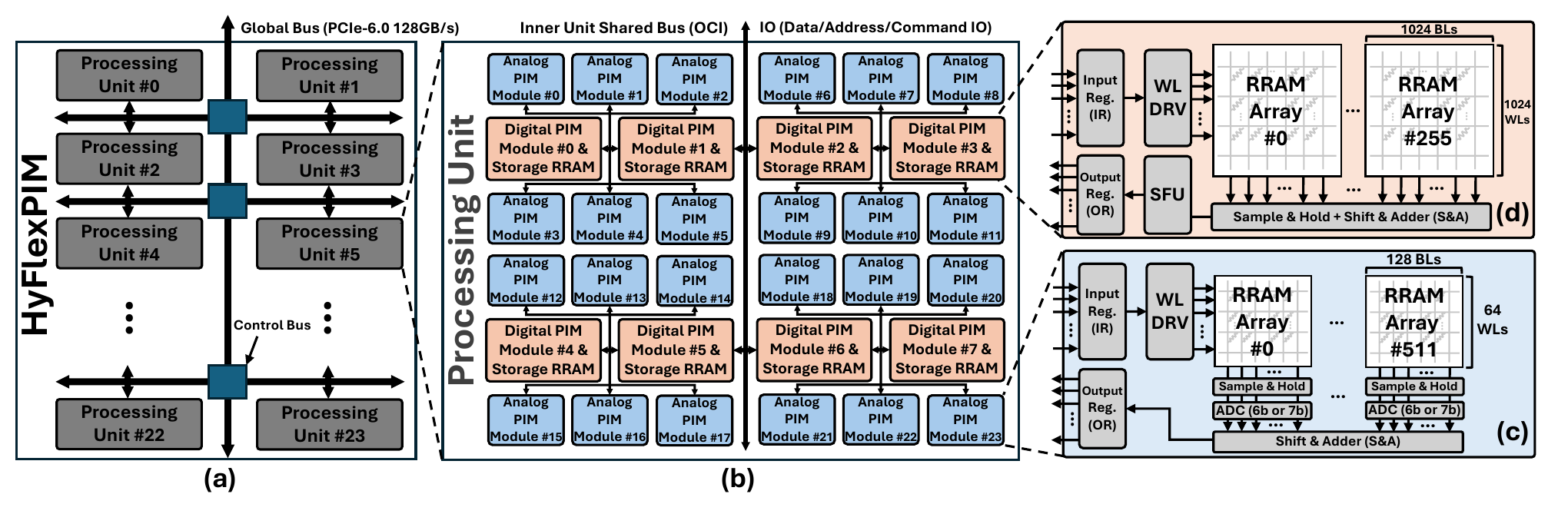}
\caption{Architecture overview of the proposed HyFlexPIM, which is based on a scalable architecture with hybrid analog and digital RRAM PIM modules.}
\label{fig:architecture}
\end{figure*} 

Digital PIM, which has been extensively studied \cite{li2024asadi, imani2019floatpim, yang2020retransformer}, achieves data movement reduction with reliable computation through simple bit-wise operations (e.g., NOR, INV, and others) but suffers from limited parallelism compared to analog PIM. While analog RRAM PIM offer superior parallelism, their non-ideality and expensive MLC write operation 
make them less suitable for storing real-time generated vectors with high bit-precision. Therefore, we opt to process attention computations in digital SLC RRAM arrays, which serve as both efficient storage and computing units.
On the other hand, the static weights in the linear layers are processed in the analog PIM for the enhanced efficiency by utilizing MLC.

As described in Figure~\ref{fig:architecture}, HyFlexPIM includes 24 Processing Units (PUs) to enable high parallelism in processing 24 layers and has a single input and output (IO) to receive input sequences and uses a global bus to communicate with each PU inside the HyFlexPIM. Each PU is dedicated to the computation of a single layer of the encoder or decoder, allowing the PUs to work in a pipelined fashion to process multiple streamed inputs. Once a specific layer is computed, the result is transferred to the next PU for the next layer's processing. Note that the output of each layer is merely the size of (\(D_h \times 1\)), maintaining the inter-PU movement cost negligible.
While mapping each PU to a single layer provides optimal performance, our architecture allows flexible model scaling in several ways as follows. 
1) For longer sequences (e.g., $>$8192) or higher hidden dimensions, multiple PUs collaboratively process a single layer to provide sufficient storage capacity, following tensor parallelism.
2) For models with fewer layers (e.g., $<$24), multiple PUs work collectively to compute a single layer in parallel, as seen in \cite{li2024asadi, liu2023hardsea, rram_isaac}, again following tensor parallelism.
3) When the number of layers exceeds the available PUs, multiple chips are cascaded to support deeper layers, following pipeline parallelism.
In cases 1) and 2), partial sums ($<$3~KB per PU) are transferred between PUs and aggregated in the Special Function Unit (SFU) via global bus (128 GB/s~\cite{sharma2020pci} as will be shown in Section~\ref{hy_config}) with negligible latency overhead (24 cycles).
For case 3), only a single hidden layer output (0.75 - 2~KB at hidden dimension of 768-2048) is transmitted between chips to continue processing the next layer, requiring less than 6-16 cycles over the PCIe-6.0 interconnect. The throughput analysis of these cases are given in Figure~\ref{fig:ratio} with real benchmarks. 
Figure \ref{fig:architecture} (b) shows the detailed structure of the PU, which comprises eight digital RRAM PIM modules for attention and non-linear operations and 24 analog RRAM PIM modules to accelerate the other linear operation stages. Each module is interconnected via  \textit{inner unit shared bus} to transfer the intermediate results from each stage to the next as input. An analog PIM module (Figure \ref{fig:architecture} (c)) consists of 512 RRAM arrays, each of which includes 64$\times$128 bitcells and peripheral circuitry such as Input/Output Registers (IR/OR), Sample and Hold (S$\&$H), 6-b/7-b flexible ADC, Shift and Adder (S$\&$A), Word Line Drivers (WL DRV) for bitwise input fetching, and conventional read circuitry and write driver (not shown for simplicity). Digital PIM module (Figure \ref{fig:architecture} (d)) is mostly similar to the analog PIM, but does not require the analog peripheral circuitry such as ADC. IR holds the control signals of bit line and WL selectors as \cite{li2024asadi, zhou2022transpim}. The Digital PIM module also incorporates a Special Function Unit (SFU) to efficiently handle non-linear operations such as Softmax, Layer Normalization, and GELU. To enable these operations, we implemented max search, subtraction, exponentiation (via Taylor series), addition, division, multiplication, and square root functions, all fully pipelined using floating-point arithmetic. After computation, the results are converted to the integer format for further processing with negligible overhead as shown in Table~\ref{tab:config}. 
We configured each SFU to process 256 inputs per cycle to balance the throughput of GEMV operations. 
The digital PIM module contains 256 RRAM arrays, each of which includes 1024 bitcell per row.
In the $\mathbf{Q} \times \mathbf{K}$ computation, which involves INT8 $\times$ INT8 multiplication, each output generation requires 64 NOR operations.  Each NOR operation takes three columns for two input operand bits and one output bit~\cite{imani2019floatpim}. Each row processing takes total five cycles (four cycles for writing and one cycle for reading). Based on these constraints, 256 inputs per cycle for SFU hit the optimal balance with GEMV throughput, e.g., 256  $\times$ 1024 $/$ (64 $\times$ 3) $/$ 5 = 273 operations.
The area and energy of SFU are in Table \ref{tab:config}.
Note, our approach is optimized for the inference, with proposed SVD fine-tuning (described in Section~\ref{SVD}) conducted prior to deployment on the hardware as a one-time software process, ensuring no additional hardware overhead for SVD processing or memory usage in inference.

\subsection{MLC vs. SLC Operations}

\begin{figure}[t]
\centering
\includegraphics[width=0.7\columnwidth]{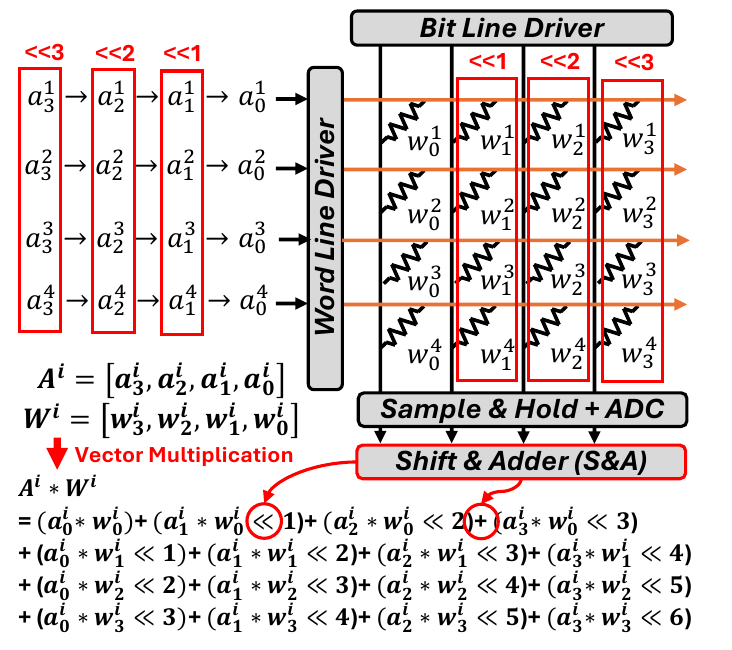}
\caption{SLC PIM mapping and Shift \& Adder operation.}
\label{fig:slc}
\end{figure} 

\begin{figure}[t]
\centering
\includegraphics[width=0.8\columnwidth]{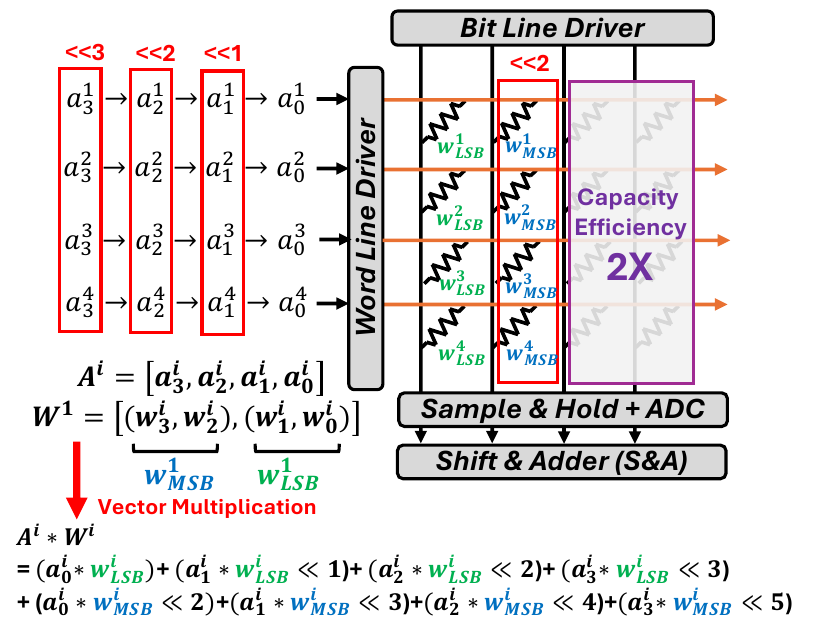}
\caption{MLC PIM mapping and Shift \& Adder operation.}
\label{fig:mlc}
\end{figure} 

The static weights are pre-loaded into the analog RRAM PIM modules before the inference begins, based on their type of storage format (e.g., SLC vs. MLC), using conventional write drivers and a wordline (WL) driver, which are omitted in Figure~\ref{fig:architecture} (c) for simplicity. Note that the varying levels of resistance for MLC are written by iteratively applying pulses through the WL based on the desired resistance level \cite{RAMADAN2019169}. As a result, the same wordline drivers are used for both SLC and MLC, incurring no additional hardware costs to support both formats.
Figure~\ref{fig:slc} and Figure~\ref{fig:mlc} show the detailed view of Figure~\ref{fig:architecture}(c) for SLC and MLC configurations to describe the PIM dot-product operation with 4-bit weights for exposition purpose. The weights in a column of the matrix are stored vertically in the RRAM array. 
The SLC stores bits for the weight across adjacent columns, e.g., 4-bit words are stored across four columns. Each bit of the input is given sequentially through the WL. Depending on the bit position of each column and the cycle of input, corresponding binary weights are applied in the digital shift~$\&$~adder (S$\&$A) module.
The MLC operation is mostly similar, but each bitcell contains multi-bits, e.g., 2-bit in Figure~\ref{fig:mlc}. Therefore, the corresponding impact factor of $1\times$, $4\times$, $16\times$, ... are given for the consecutive columns in the  S$\&$A module. Note that the support of SLC and MLC does not require two distinct hardware, but sharing the same hardware in a reconfigurable manner with only minor additional functions in the digital S$\&$A module and reconfigurable ADC.

\begin{figure}[t]
\centering
\includegraphics[width=0.8\columnwidth]{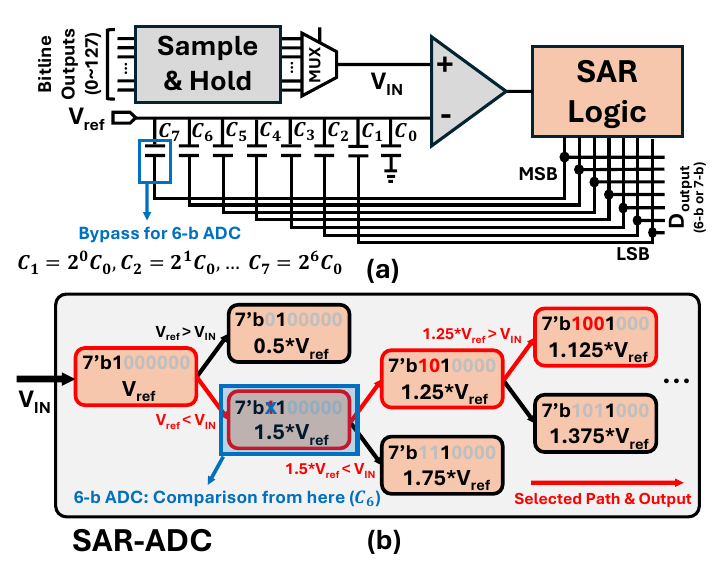} 
\caption{(a) 6 and 7 bits flexible SAR-ADC schematic, and (b) operation of SAR-ADC with reconfiguration.} 
\label{fig:ADC}
\end{figure}

By Kirchhoff's current law, the currents accumulated by bitlines represent the dot-product results in the analog domain. As described in Figure \ref{fig:ADC} (a), analog signals from 128 bitlines are first captured by sample and hold (S\&H) circuits, then sequentially selected through a multiplexer before being fed to a shared successive-approximation-register (SAR) ADC. The SAR ADC performs analog-to-digital conversion using binary search algorithm by enabling capacitors in sequence ($\mathbf{C_7}$ → $\mathbf{C_1}$) for iterative voltage comparisons, where each comparison determines one bit of the digital output. The required ADC precision for MVM operations is given by $\left\lceil {\log_2(R)} \right\rceil  + w - 1$, where $R$ is the number of rows in the crossbar, $w$ is the bits stored in a bitcell.
In the SLC case,  an 6-bit ADC is needed to achieve the full bit precision assuming 64 rows. 
In the MLC PIM operation, as shown in Figure~\ref{fig:mlc}, two consecutive bits of weights are packed and mapped to a single 2-bit RRAM device. In this way, MLC improves efficiency with doubled throughput, increased storage density, and halved analog computation energy. Since we still send input bits one by one, the ADC resolution required for MLC only increases by one (6-b → 7-b) compared to that of the SLC operation. While numerous studies have explored truncating bit precision of ADC to reduce area and energy consumption \cite{li2024asadi, liu2023hardsea}, Transformers are highly sensitive to such a bit truncation, leading to significant accuracy degradation. 
To mitigate accuracy degradation, we employ a full-precision ADC, configured as 6-b for SLC and 7-b for MLC, in a flexible bit-precision manner. As shown in Figure \ref{fig:ADC} (b), our ADC is designed to support up to 7-b resolution. However, by simply bypassing the comparison for the largest capacitor (\( \mathbf{C_7} \)), which corresponds to the MSB, the ADC can efficiently operate as a 6-b ADC without consuming additional power. Importantly, the energy and area overhead of this design compared to a dedicated 6-b ADC takes negligible (<1\%) portion in the overall architecture. Generally, increasing the ADC precision by one bit results in doubling the energy consumption \cite{kull20133, adc_survey}. However, because the number of results generated to be converted by ADC is reduced by half in the case of 2-b MLC, the total energy consumption for the ADC does not increase as compared to the SLC case. Based on reliability considerations, we adopt 2-b MLC to balance efficiency and reliability since measurements from a real RRAM chip \cite{wan202033, fan2024efficientopenmodificationspectral} show that higher-level MLCs (e.g., 3-b/4-b MLC) have a 7× higher bit error rate than SLC, causing severe accuracy degradation in Transformer models \cite{devlin2018bert, radford2019language}.

\subsection{Transformer Algorithm to HW Mapping}

\begin{figure}[t]
\centering
\includegraphics[width=0.9\columnwidth]{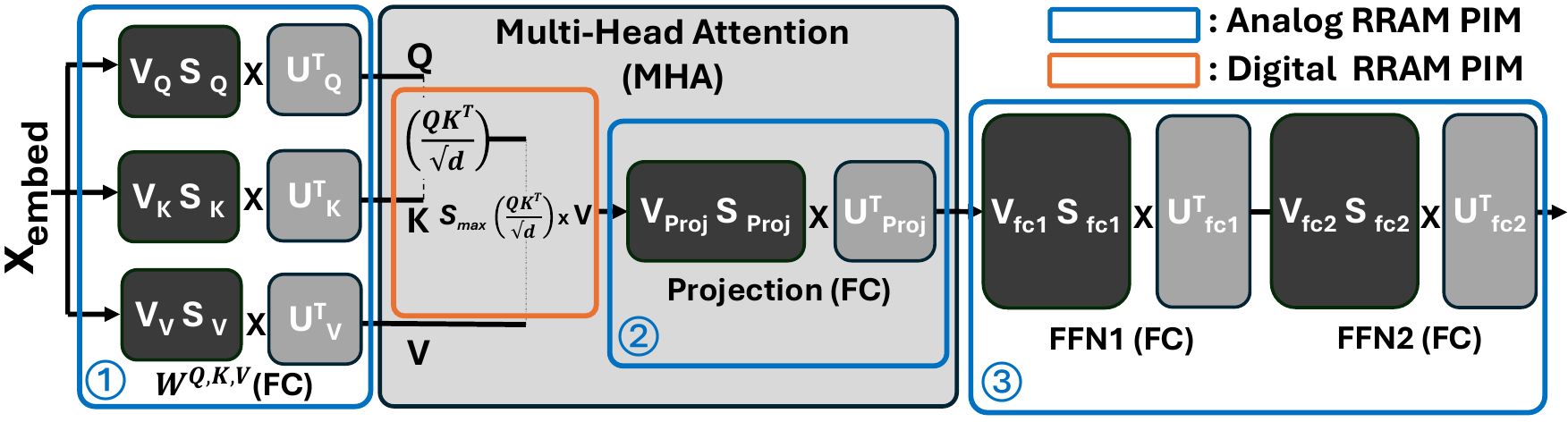} 
\caption{The proposed mapping of transformer operations to the hybrid digital-analog RRAM PIM architecture.} 
\label{fig:dataflow}
\end{figure}

Figure \ref{fig:dataflow} illustrates the dataflow of Transformer including the four fully-connected (FC) layers, and MHA layer and their mapping on PIM. After applying SVD, truncation on the linear matrices $\mathbf{W^Q}$, $\mathbf{W^K}$, and $\mathbf{W^V}$, and fine-tuning, the matrices are divided into two matrices: $V$ $\times$ $\Sigma$ and $U^T$ (Since input is multiplied with transposed version of $\mathbf{W}$) and stored in the analog PIM RRAM array for the inference as shown in Figure \ref{fig:dataflow} block \circled{1}. After matrix multiplication with an input  ($X_{embed}$) with these two matrices, $\mathbf{Q}$, $\mathbf{K}$, and $\mathbf{V}$ tokens are generated. The generated tokens or outputs are stored in the digital PIM module via the inner-unit shared bus, where partial sums from different sub-arrays are accumulated to finalize the GEMV computation before passing the results to the next stage. While recent analog RRAM PIM works tend to employ the SLC due to the non-ideality concerns  \cite{liu2023hardsea, li2024asadi}, our approach utilizes more efficient MLCs for the most (e.g., 90 - 95\% for BERT) of less critical elements by relying on the 
proposed gradient redistribution with minimal impact ($<$1\%) on accuracy as detailed in Section~\ref{sec:accuracy}. The same process is applied on FC layer \circled{2} in the MHA and two FFN layers \circled{3} by storing their static weights in the RRAM array and performing the PIM operation by applying the output of previous layer as an input.

While the weights of the above layers are static, the computations of $\mathbf{Q} \cdot \mathbf{K^T}$ and $\times \mathbf{V}$ (in the orange box) are not static, i.e., both operands are computed in real-time given input. Therefore, we do not apply gradient redistribution to avoid a costly SVD computation on the fly, in contrast to static weights where SVD and gradient redistribution are pre-computed offline. We also employ only SLC to avoid the costly MLC writing operation, which requires iterative verify-read and write operations to reach the exact target conductance. Furthermore, this stage has relatively lower computation volume and requires high precision computations, e.g.,  $\times \mathbf{V}$ stage requires an operand to be  12 bits, to guarantee accuracy for a wide range of applications. For this reason, we employ digital PIM for this section (orange box) as proposed in \cite{li2024asadi, imani2019floatpim, 9663036, 10.1145/3400302.3415640} to avoid any potential accuracy degradation and excessive ADC costs. Although the benefit of digital PIM is limited compared to analog PIM, it still bypasses the expensive data movement cost for the KV cache from the memory to the processor. We also store the intermediate computed results in the digital PIM with an SLC format.

As described in Figure~\ref{fig:transformer},  there are two distinct types of Transformer tasks: the encoder and decoder stages. In the encoder (or the prefill phase of the decoder), multiple input vectors are applied to PIM through the wordlines  over multiple cycles to perform matrix-matrix multiplications. In contrast, during the decoder phase, a single input vector is applied for vector-matrix multiplication, generating one token at a time. Nevertheless, the PIM operations remain the same, ensuring consistency in execution. 

\section{SVD-based Gradient Redistribution} \label{SVD}

This section introduces the SVD-based \textit{gradient redistribution} technique, designed to maximize the efficiency of the HyFlexPIM accelerator. By applying SVD and truncation, we expose the importance of weight elements and concentrate critical weights into a smaller subset through fine-tuning. This hardware-algorithm co-design preserves critical weights in SLC while utilizing efficient MLC RRAM for the majority of less-critical weights, maximizing the HyFlexPIM architecture efficiency.

\begin{figure}[t!]
\centering
\includegraphics[width=1\columnwidth]{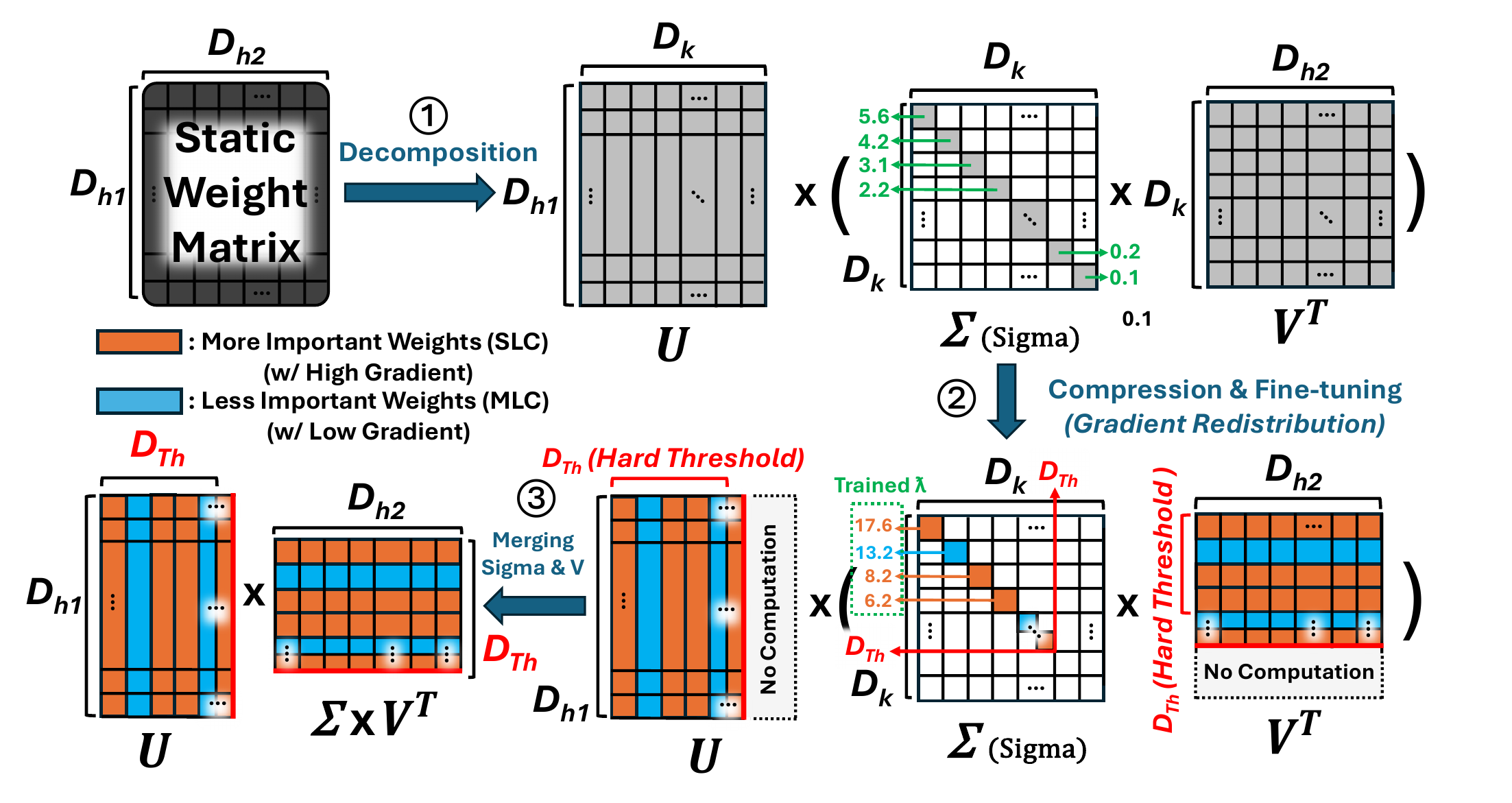}
\caption{Hard-Thresholding and Fine-tuning the post-SVD matrix.}
\label{fig:numcomputation}
\end{figure} 

\begin{figure*} [h]
\centering
\includegraphics[width=1.8\columnwidth]{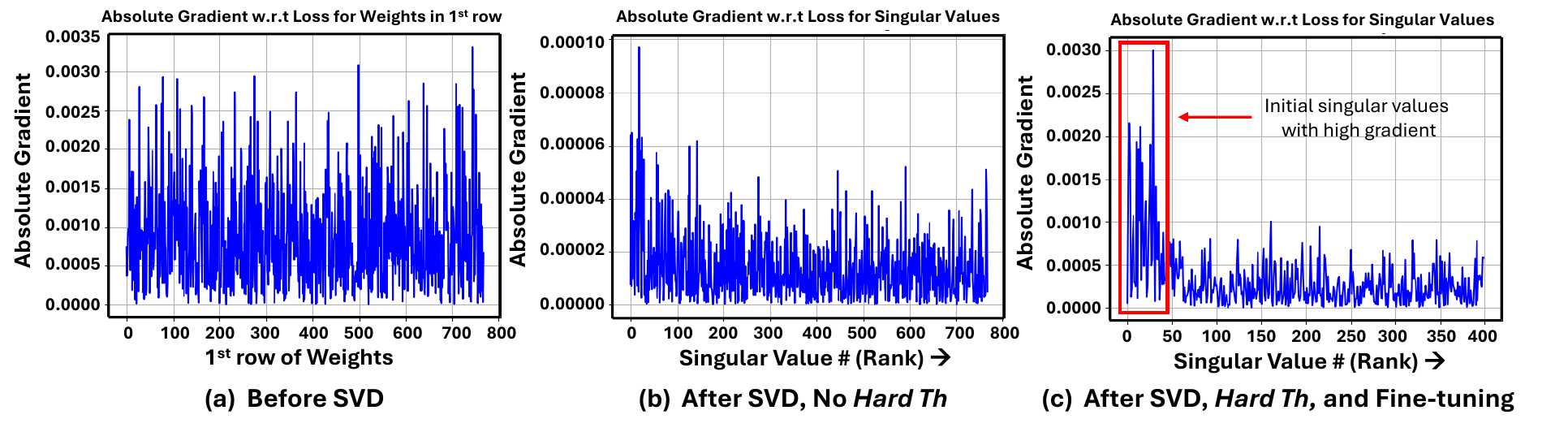}
\caption{Gradient distribution change for FC layer in BERT-Base: (a) gradient of each weight (order-wise) in a row before SVD, (b) gradient after SVD, without applying hard threshold, (c) gradient after SVD, applying hard threshold and gradient redistribution.}
\label{fig:finetuning}
\end{figure*} 

\subsection{Post-SVD Truncation}

\begin{algorithm}
\caption{Hybrid SLC-MLC RRAM Analog PIM Simulation}
\label{alg:hybrid}
\begin{algorithmic}[1]
\Require{Weight matrix \( W \in \mathbb{R}^{m \times n} \), desired rank \( k \), and optimizer (AdamW)}
\Ensure{Optimized weight mapping for HyFlexPIM}
\State \textbf{Step 1: SVD Decomposition}
\State Perform SVD on \( W \): \( W = U \Sigma V^\top \)
\State \textbf{Step 2: Truncated SVD}
\State Truncate matrices: \( U \in \mathbb{R}^{m \times k} \), \( \Sigma \in \mathbb{R}^{k \times k} \), \( V \in \mathbb{R}^{k \times n} \)
\State \textbf{Step 3: Fine-tuning}
\While{training \textbf{for} 1-3 epochs}
  \State Perform optimizer step (e.g., AdamW)
  \State Store gradients and update weights
\EndWhile
\State \textbf{Step 4: Gradient-based Rank Selection}
\State Identify top singular values
\State \textbf{Step 5: Mapping Weights into PIM array}
\State Map critical weights to SLC, others to MLC
\end{algorithmic}
\end{algorithm}

As mentioned in Section~\ref{SVD_background}, by applying SVD, the importance of weight data becomes more discrete due to the ordered singular values in \( \Sigma \). This facilitates mapping critical data to SLC RRAM, enhancing robustness and accuracy, while less critical data can be mapped to MLC RRAM, optimizing area, latency, and energy efficiency. However, direct application of SVD increases computational overhead due to the additional matrix multiplications. To address this issue, we employ a hard-threshold for \( \Sigma \) that maintains the same computational complexity as the pre-SVD matrices. Figure \ref{fig:numcomputation}, step \circled{2}, illustrates the computation reduction process. For a weight matrix \(W\) of size \(D_{h1} \times D_{h2}\), the total number of MAC computations before SVD is \(L \times D_{h2} \times D_{h1}\), and the total number of parameters is \(D_{h1} \times D_{h2}\). We truncate the \( \Sigma \)  to keep the limited number of ranks. 
Then, we combine \( \Sigma \) and \( V^T\), e.g., \( \Sigma \ \times V^T\), to reduce the number of parameters for the inference, as described in Figure~\ref{fig:numcomputation}, step \circled{3}. Consequently, the number of MAC operations changes to \(L \times D_{h2} \times D_{Th} + L \times D_{Th} \times D_{h1}\), where \(D_{Th}\) is the lower rank. To ensure that the computational load remains the same as the original model, we replace \(D_{Th}\) with a hard threshold \(\frac{D_{h1} \times D_{h2}}{D_{h1} + D_{h2}}\).
By using the hard threshold (\(\frac{D_{h1} \times D_{h2}}{D_{h1} + D_{h2}}\)), the total number of parameters for \( \Sigma \times V^T \) and \( U \) is also preserved before and after SVD.

The low-rank truncation  causes some accuracy degradation. For example, in BERT-Large, the MRPC task shows a 16\% accuracy drop, RTE shows a 20\% drop, and in GPT-2, the loss increases by more than 2$\times$. To recover accuracy after truncation, we fine-tune the model by re-training for 1-3 epochs, as described in \textbf{Steps 1-3} of Algorithm~1, which we found sufficient to maintain accuracy and loss comparable to the pre-SVD case for transformer models used in our evaluation. Simpler tasks often required only a single epoch of fine-tuning, effectively balancing computational efficiency with parameter reduction. Moreover, the SVD decomposition, truncation, and fine-tuning (\textbf{Steps 1 - 4} in Algorithm 1) are performed entirely in software, while the final matrices \(V \times \Sigma\) and \(U^T\) are offloaded to the hardware for GEMV operations. Thus, the fine-tuning process incurs only one-time cost without adding any hardware overhead.

\subsection{Gradient Redistribution}
\label{sec: gradient-redistribution}

While the fine-tuning in \textbf{Step 3} aims to recover the accuracy, it also performs a crucial role, \textit{gradient redistribution}, which helps to increase the portion of  error-tolerant  parts dramatically, facilitating the use of MLC for most of computation.
The magnitude of the gradient of the loss function with respect to a specific weight in a neural network provides vital insights into the importance of the weight.
In addition, the gradient of the loss function $\mathcal{L}$ with respect to a weight $w_i$, denoted $\frac{\partial \mathcal{L}}{\partial w_i}$, signifies how sensitive the loss is to changes in the weight. A large gradient implies that a small change in the weight will result in a significant change in the loss, thus highlighting the weight's critical importance to the model's performance.
During training, weights are updated in the direction that minimizes the loss function. 

Figure \ref{fig:finetuning} shows the gradient redistribution resulting from the fine-tuning process. Since there is no notion of rank before applying SVD, we examined the gradients of the weights in a row in Figure \ref{fig:finetuning} (a). The gradients across all weights are rather uniformly distributed, making it challenging to identify which ranks are more sensitive to loss. After applying SVD (Figure \ref{fig:finetuning} (b), which is not hard-thresholded yet), the differences in gradients among the ranks remain insufficiently distinct to clearly indicate their importance. However, by fine-tuning the weights as shown in Algorithm 1 \textbf{Step 3}, in the post-SVD matrix, the gradients are re-distributed towards the higher ranks, resulting in only a small portion of ranks having significantly higher magnitude. As illustrated in Figure 
\ref{fig:finetuning} (c), the initial singular values now exhibit much higher gradients compared to the lower ranks, clearly indicating which ranks should be assigned to SLC for precise computation (Algorithm 1 \textbf{Step 4}). We attribute this redistribution to the fine-tuning process that attempts to recover the loss of information from the truncated ranks by putting more information on the untruncated ranks. In this process, the ranks with higher singular values tend to gain more information  than the others, as these ranks are principal key components to represent the matrix. 

Note above nature helps not only the clear demarcation between error-resilient (on MLC) and error-tolerant parts (on SLC), but also facilitate more computations in MLC, dramatically improving the energy, throughput, and area efficiencies.
By storing the gradient distribution after fine-tuning the truncated decomposition, we analyze the gradients for the singular values of the $\Sigma$ matrix. We choose singular values with the top k\% (k $ = $ 0, 5, 10, 30, 40, 50, 100) gradient magnitudes to perform the computation on SLC, while the rest is computed using MLC (Algorithm 1 \textbf{Step 5}). In this manner, we use a hybrid approach to minimize the accuracy degradation and maximize efficiency. We study the selection of optimal $k$ further in Section~\ref{sec:eval}.

\section{Methodology}

\subsection{Software Benchmarks}
\begin{table}[h]
  \caption{Hyper-parameters for fine-tuning.}
  \centering
  \begin{tabular}{lccc}
    \toprule
    \textbf{Model} & \textbf{Batch Size} & \textbf{Learning Rate} & \textbf{Optimizer} \\
    \midrule
    BERT-Base & 32 & 2e-5 &  \\
    BERT-Large & 32 & 5e-6 &  \\
    GPT-2 & 2 & 2e-5 & AdamW \\
    Llama3 & 2 & 2e-5 &  \\
    ViT-Base & 10 & 5e-6 &  \\
    \bottomrule
  \end{tabular}
  \label{tab:finetune_hparam}
\end{table}

The proposed HyFlexPIM is evaluated on various Transformer-based models for both NLP and CV domains including the encoder models BERT-Base and BERT-Large, the decoder model GPT-2-Small (GPT-2) and Llama-3.2-1B (Llama3), and the vision transformer model (ViT-Base) for object detection. For the encoder models, 7 datasets from the General Language Understanding Evaluation (GLUE) benchmark \cite{wang2018glue}, which includes cola, mrpc, qnli, qqp, rte, sst-2, and sts-b, are used for the evaluation. The maximum sequence length (MSL) for all GLUE datasets is set to 128. The decoder model GPT-2 is tested on the WikiText-2 dataset \cite{merity2016pointer} with an MSL of 1024, and Llama3 is tested on PTB~\cite{marcus-etal-1993-building} with an MSL of 100. ViT-Base is evaluated on the CIFAR-10 dataset with an input resolution of 224$\times$224 and a patch size of 16$\times$16. Throughout this work, we employ INT8 for all the linear layers and for $\mathbf{Q}$, $\mathbf{K}$, and $\mathbf{V}$ in the attention mechanism, while FP16 is used for other non-linear operations which are computed by HyFlexPIM’s SFU. Table~\ref{tab:finetune_hparam} shows the batch size, learning rate, and optimizer used for fine-tuning Transformer models. All the fine-tuning tasks were conducted on a server with a 12-core Intel Xeon Silver 4214R CPU and two NVIDIA GeForce RTX A6000 GPUs.

\subsection{RRAM PIM Non-ideality Considerations} \label{rram_simulator}

We developed a simulator to evaluate the non-idealities of analog behavior in SLC-MLC  PIM  modules along with other accuracy impacts from SVD truncation, fine-tuning, and the distribution of analog and digital PIM modules. The simulator also models the serial bit computing mechanism and shift-and-add operations to accurately capture the accuracy impact of this specific operation order. 
This simulation incorporates the noisy RRAM behavior with non-ideal resistance deviating from the target. Specifically, we injected noise into the parameters of all linear layers, following the formula:
\begin{equation}
\tilde W = W \odot (1 + \eta ),
\label{eq: noise}
\end{equation}
where $\eta$ is a Gaussian noise tensor with the same size as the weight tensor $W$, $\odot$ is element-wise multiplication, $\tilde W$ is the noisy weight tensor used for inference. The noise level was derived from the studies by Wan et al. \cite{wan202033} and Fan et al. \cite{fan2024efficientopenmodificationspectral}, utilizing the non-idealities measured from the fabricated real MLC RRAM chips. Using the experimental data from 3 million RRAM cells in \cite{fan2024efficientopenmodificationspectral}, which showed a bit error rate (BER) of approximately 4.04\% after one day of programming, we reverse-calculated the standard deviation of the Gaussian noise $\eta$ in (\ref{eq: noise})
to match the BER of real analog RRAM devices. On top of the RRAM device noise, we also considered the limited ADC resolution for the outputs. 

Analog RRAM module processes static weights, enabling multiple inferences after a single write operation and thus being free from endurance concerns. For the digital PIM module, where real-time generated $\mathbf{Q}$, $\mathbf{K}$, $\mathbf{V}$, and intermediate data are supposed to be written, could potentially compromise RRAM endurance \cite{yang2020retransformer}.
However, given large amount of RRAM capacity in HyFlexPIM and typical RRAM endurance of $10^8$ cycles \cite{8640245}, HyFlexPIM ensures sustainable operation beyond typical server lifespans (3-5 years) even with 10K daily inference requests.

\subsection{Hardware Baseline} \label{baseline}

We benchmark our HyFlexPIM against 1) the latest study in analog-digital hybrid PIM architecture for Transformer models ASADI~\cite{li2024asadi} using FP32, which only uses SLC PIM to contrast our effort to use the MLC for most computations; 2) ASADI$^\dagger$, using INT8 precision for linear layers which provides conservative comparison by improving ASADI's efficiency; 3) SPRINT~\cite{yazdanbakhsh2022sparse}, which utilizes the digital processor with RRAM storage for the linear layer computation while using analog PIM for the limited part of attention (MSB 4-bit computation of $\mathbf{Q} \cdot \mathbf{K}$); 4) a near-memory processing (NMP) baseline, TransPIM~\cite{zhou2022transpim}, which is based on the  function-in-memory DRAM (FIMDRAM) processor~\cite{kwon202125} with HBM; and 5) a non-PIM baseline that processes data in a digital module consisting of INT8 dot product units derived from SPRINT~\cite{yazdanbakhsh2022sparse}, after moving data from 6.28~GB off-chip DRAM to on-chip SRAM cache, assuming large enough cache capacity.
For direct comparison, all results from these baselines are scaled to the same 65nm process technology, following the methodology in \cite{stillmaker2017scaling}. For energy estimation, we utilize NVSIM~\cite{6218223} for RRAM, ARM Memory Compiler with high-density 65~nm single-port SRAM (version r0p0) \cite{mem_compiler} with 1~GHz operating frequency, and DRAM access energy from \cite{wang2024beacongnn}. The near-memory processing bank energy is from~\cite{zhou2022transpim}  while HBM-related data is sourced from~\cite{o2017fine}. We evaluate ASADI and SPRINT's methodologies within our hardware to assess their effectiveness in our optimized mixed-signal PIM architecture.  

\begin{table}[t]
\setlength{\tabcolsep}{3pt}
\caption{The hardware configuration and component-level area, power of HyFlexPIM.} 
\label{tab:config}
\small
\begin{tabular}{|c|c|c|c|c|}
\hline
\textbf{Component} & \textbf{Area ($mm^2$)} & \textbf{Power (mW)} & \textbf{Parameter} & \textbf{Value} \\
\hline
\multicolumn{5}{|c|}{\textbf{Analog RRAM Module}} \\
\hline
\multirow{3}{*}{\textbf{RRAM Array}} 
& \multirow{3}{*}{0.048 (9.6\%)} & \multirow{3}{*}{60.78 (6.5\%)} 
& Bit/Cell & 1-b/2-b \\
& & & Size & 64×128 \\
& & & Total & 512 \\
\hline
\textbf{IR} & 0.00065 (0.1\%) & 0.13 (0.01\%) & Size/each & 64B \\
\hline
\textbf{OR} & 0.00129 (0.3\%) & 0.53 (0.1\%) & Size/each & 128B \\
\hline
\multirow{2}{*}{\textbf{WL DRV}} 
& \multirow{2}{*}{0.02 (3.7\%)} & \multirow{2}{*}{297.71 (32\%)} 
& Resolution & 1-b \\
& & & Total & 64×512 \\
\hline
\multirow{2}{*}{\textbf{ADC}} 
& \multirow{2}{*}{0.30 (64.2\%)} & \multirow{2}{*}{512.00 (55\%)} 
& Resolution & 6-b/7-b \\
& & & Total & 512 \\
\hline
\textbf{S\&A} & 0.10 (22\%) & 59.54 (6.4\%) & Total & 512 \\
\hline
\textbf{S\&H} & 6e-5 (0.01\%) & 12e-6 (0.001\%) & Total & 512 \\
\hline
\textbf{Sum} & 0.47 & 930.69 & - & - \\
\hline
\textbf{Total} & \textbf{11.24} & \textbf{22,336.59} & - & 24 \\
\hline
\multicolumn{5}{|c|}{\textbf{Digital RRAM Module}} \\
\hline
\multirow{3}{*}{\textbf{RRAM Array}} 
& \multirow{3}{*}{2.86 (35.8\%)} & \multirow{3}{*}{3,890.02 (59.6\%)} 
& Bit/Cell & 1-b \\
& & & Size & 1024×1024 \\
& & & Total & 256 \\
\hline
\textbf{IR} & 0.0031 (0.04\%) & 0.76 (0.01\%) & Size/each & 1KB \\
\hline
\textbf{OR} & 0.0032 (0.04\%) & 1.65 (0.03\%) & Size/each & 1KB \\
\hline
\multirow{2}{*}{\textbf{WL DRV}} 
& \multirow{2}{*}{0.14 (1.8\%)} & \multirow{2}{*}{2,381.64 (36.5\%)} 
& Resolution & 1-b \\
& & & Total & 1024×256 \\
\hline
\textbf{S\&A} & 0.21 (2.6\%) & 119.08 (1.8\%) & Total & 1024 \\
\hline
\textbf{S\&H} & 13e-5 (0.002\%) & 23e-6 (0.0\%) & Total & 1024 \\
\hline
\textbf{SFU} & 4.79 (59.8\%) & 138.89 (2.1\%) & \# inputs & 256 \\
\hline
\textbf{Sum} & 8.01 & 6,532.05 & - & - \\
\hline
\textbf{Total} & \textbf{64.05} & \textbf{52,256.41} & - & 8 \\
\hline
\end{tabular}
\end{table}

\begin{figure*}[t]
\centering
\includegraphics[width=2\columnwidth]{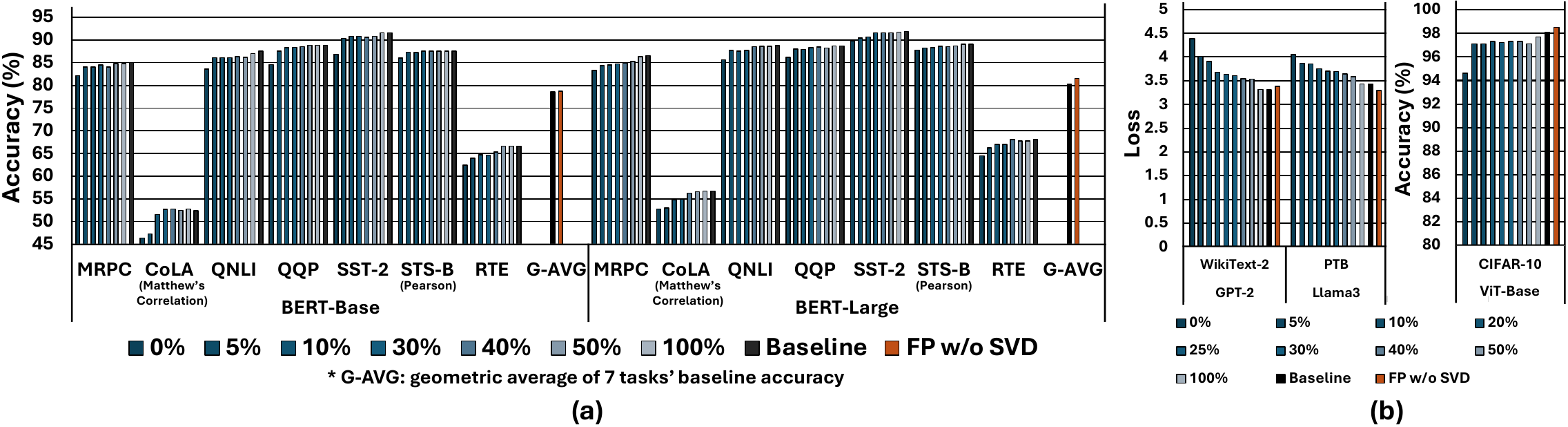}
\caption{Accuracy trends with respect to the SLC rate for (a) BERT-Base, BERT-Large, (b) GPT-2, Llama3, and ViT.}
\label{fig:bert_accuracy}
\end{figure*} 

\subsection{HyFlexPIM Configuration}  \label{hy_config}

HyFlexPIM is designed to store all original and intermediate data within RRAM arrays. 
This also makes HyFlexPIM's memory capacity be dependent on the input sequence length. The HyFlexPIM is configured to handle the MSL of 8192. For BERT-Large has 24 layers, with each PU assigned to process a single layer. On the other hand, BERT-Base and GPT-2, having 12 layers, benefits from a 2× throughput increase with our accelerator.
The analog PIM module comprises 512 arrays of 64×128 (1KB) RRAM for INT8 $\mathbf{W^Q}$, $\mathbf{W^K}$, and $\mathbf{W^V}$, as well as Proj (FC), FFN1, and FFN2 matrices. Meanwhile, the digital PIM module includes 256 arrays of 1024×1024 (128KB) RRAM for intermediate matrices.

We adopt the data transfer and on-chip communication topologies in \cite{saberi2011analysis}, which also used full-flow RRAM. Inter-PU data transfer is facilitated by a 1000~GB/s On-Chip Interconnect (OCI) \cite{jouppi2021ten} and a PCIe-6.0 interface with 128~GB/s bandwidth for outer-PU communication \cite{sharma2020pci}. Detailed specifications of the RRAM arrays are provided in Table~\ref{tab:config}. The analog modules utilize 1-bit (SLC) or 2-bit (MLC) one transistor and one memristor (1T1M) RRAM cells, while the digital modules use only 1-bit 1T1M RRAM arrays. Area and power configurations for the 1T1M RRAM arrays are derived from \cite{wan202033, li2024asadi}. RRAM arrays read/write data in a column-parallel manner, with SET/RESET voltages of 1.62/3.63~V for 1-bit cells \cite{hung2021four}. Specifically, the RRAM features an on-state resistance (\(R_{\text{ON}}\)) of 6~k\(\Omega\) with an on/off ratio of 150 \cite{yu2021rram}. To manage 6-bit/7-bit data across 64 rows using 1-bit/2-bit cells, a shared 6-bit/7-bit reconfigurable ADC is employed for 128 columns.
The bitline currents are captured in 128 sample-and-hold circuits \cite{10.1145/3007787.3001139, kull20133}. During the following 100~ns cycle, these analog values are sequentially input into a single 1.28 GSps ADC unit, while the crossbar initiates its next read operation in a pipelined manner. Consequently, 128 bitline currents are converted to digital within 100~ns before latching the subsequent set.

The area and power specifications for the S\&A unit and on-chip SRAM buffers (IR and OR) are sourced from \cite{shafiee2016isaac, li2024asadi}, with the DRV details obtained from a 1-bit digital-analog converter (DAC) in \cite{saberi2011analysis, li2024asadi}. We validate the behavior of RRAM arrays using a modified NVSIM \cite{6218223}. Additionally, we developed an error-noise injected simulator to evaluate the accuracy of HyFlexPIM across various models and datasets as described in Section~\ref{rram_simulator}. All values have been scaled to a 65~nm technology node based on \cite{stillmaker2017scaling}. We use a 65 nm general-purpose standard cell library to synthesize SFU at 1~GHz operating frequency with Cadence Genus 19.1.




\begin{figure}[t]
\centering
\includegraphics[width=1\columnwidth]{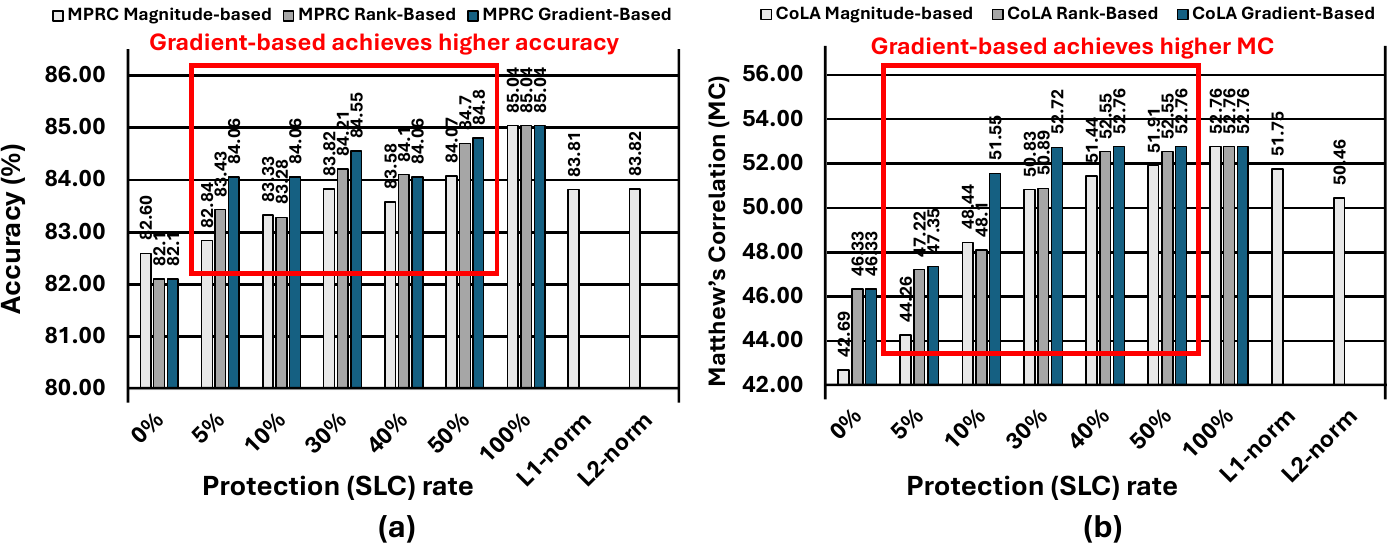}
\caption{Accuracy comparison between weight magnitude-based vs. rank-based after SVD vs. gradient-based after SVD SLC selections in BERT-Base with (a) MRPC and (b) CoLA datasets.}
\label{fig:protection}
\end{figure}

\begin{figure*}[th]
\centering
\includegraphics[width=2\columnwidth]{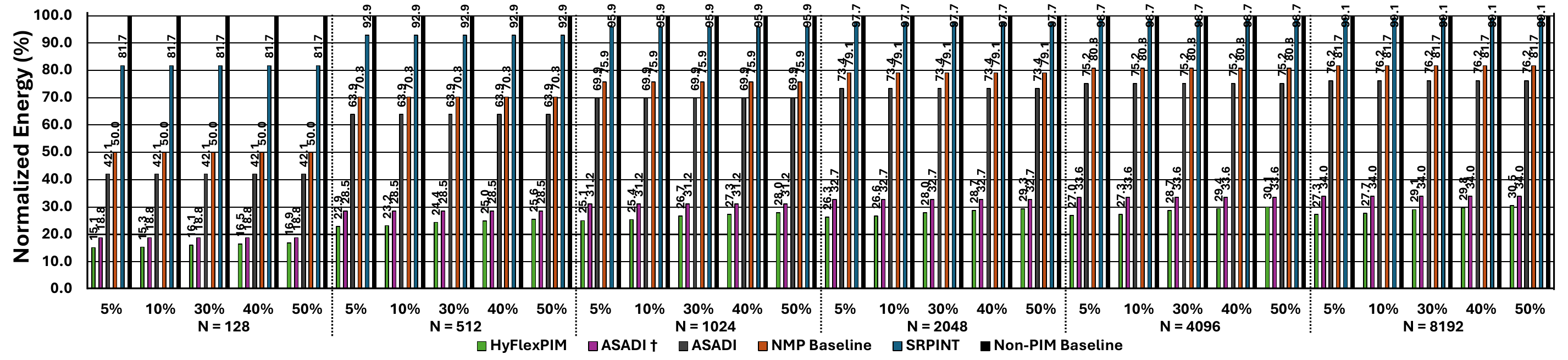}
\caption{Normalized energy consumption of linear layers compared to the energy of ASADI$^\dagger$, ASADI~\cite{li2024asadi}, NMP baseline, SPRINT~\cite{yazdanbakhsh2022sparse}, and non-PIM baseline.}
\label{fig:energy}
\end{figure*} 

\begin{figure*}[th]
\centering
\includegraphics[width=2\columnwidth]{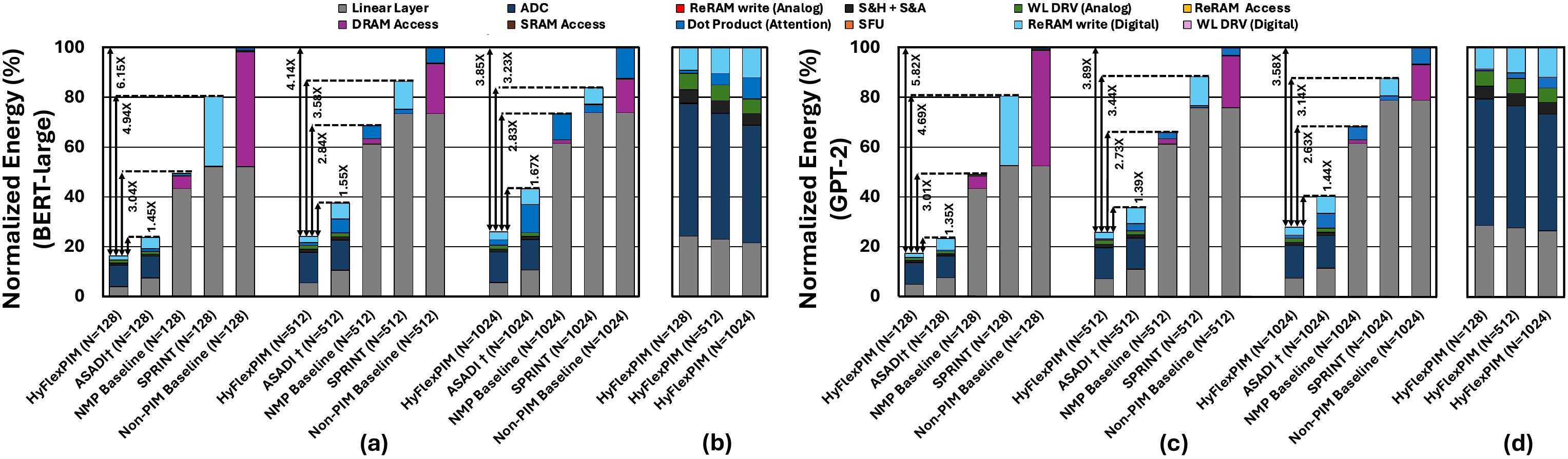}
\caption{End-to-end energy: (a) comparison with SOA, and (b) breakdown with HyFlexPIM at 5\% SLC for BERT-Large, and  (c) comparison with SOA, and (d) breakdown with HyFlexPIM at 30\% SLC for GPT-2.}
\label{fig:energy_breakdown}
\end{figure*} 

\begin{figure}[th]
\centering
\includegraphics[width=0.9\columnwidth]{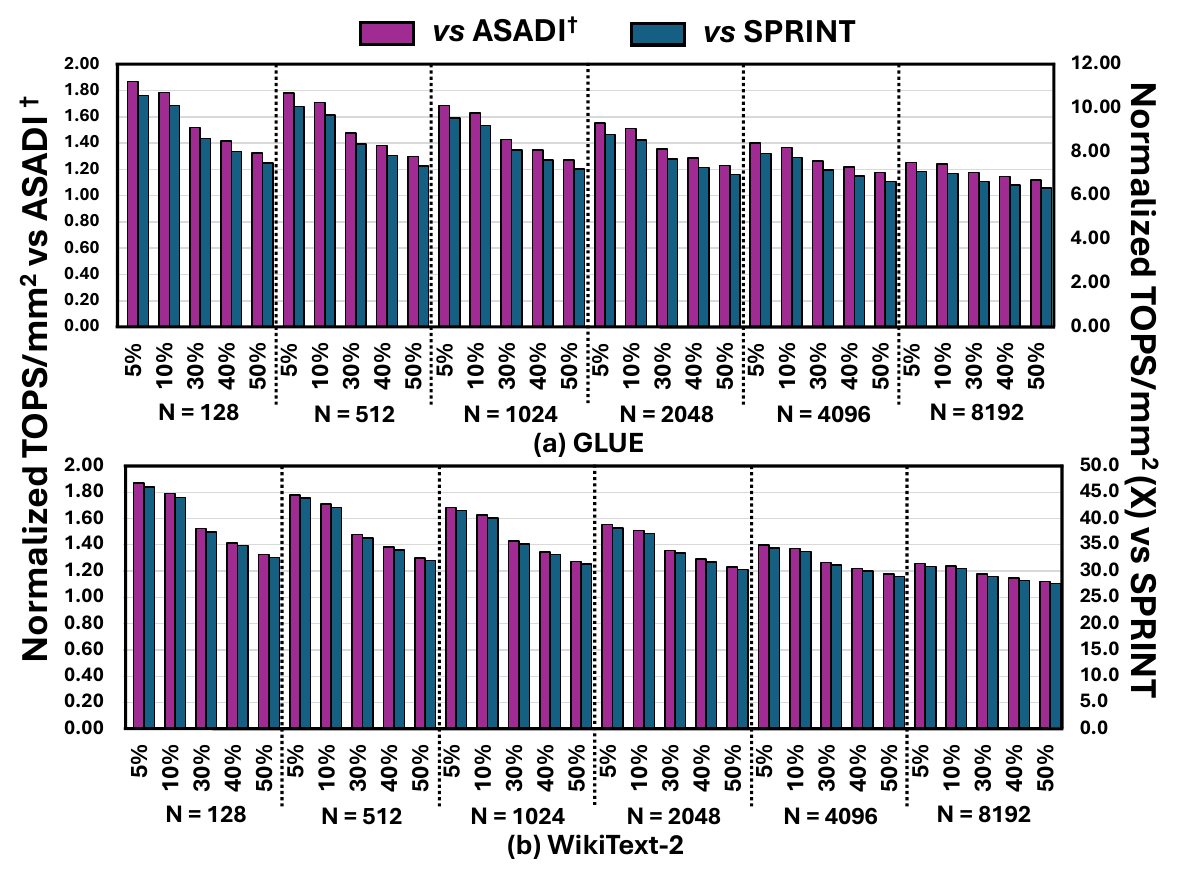} 
\caption{Speedup factor compared to ASADI$^\dagger$ \cite{li2024asadi} and SPRINT \cite{yazdanbakhsh2022sparse} with respect to the SLC rate and the sequence length.}
\label{fig:speed}
\end{figure}


\section{Evaluation}\label{sec:eval}

\subsection{Accuracy Recovery by Mapping on SLC} \label{sec:accuracy}
Figure \ref{fig:bert_accuracy} illustrates the accuracy of each transformer models resulting from the hybrid SLC-MLC RRAM mapping technique with different protection rates. k\% protection rate implies that weights corresponding to the singular values with top k\% of the absolute gradients will be mapped on SLC, while the rest will be mapped on MLC. The baseline represents the accuracy with INT8 precision 
without in-memory computing noise, but with SVD decomposition and fine-tuning. FP w/o SVD refers to models that use floating-point without applying SVD decomposition. The accuracies of FP w/o SVD  for BERT are from~\cite{devlin2018bert}, GPT-2 from~\cite{Radford2019LanguageMA}, and ViT from~\cite{yoshioka2024visiontransformers}, while we fine-tuned Llama3. For the BERT-Base (Figure \ref{fig:bert_accuracy} (a)), it is observed that maintaining a 5\% SLC rate in MRPC, QNLI, and STS-B tasks minimizes accuracy degradation to less than 1\% compared to the baseline model with quantization. In CoLA, QQP, SST-2, and RTE by increasing the SLC rate to 10-30\%, the accuracy drop is maintained to be less than 1\% of the baseline. In the case of BERT-Large, the QNLI, QQP, STS-B, and RTE tasks require only a 5\% SLC rate while the others maintain the accuracy with 10-30\% rate, confirming greater robustness against errors from MLC compared to smaller models. In Figure \ref{fig:bert_accuracy} (b), we assessed the evaluation loss on the WikiText-2 and PTB dataset for GPT-2 and Llama3 models, respectively. With an SLC rate of 20\%, we observe less than a 10\% increase in loss compared to the case with a 100\% SLC rate. 
Furthermore, for the ViT-Base on the CIFAR-10 dataset, we observe that an SLC rate of only 5\% shows less than 1\% accuracy drop. These findings indicate that strategic selection of SLC rates is crucial for maintaining high accuracy across different tasks and model sizes.

\subsection{Gradient vs. Rank vs. Magnitude based SLC mapping} \label{sec:selecting_SLC}

We further investigated the advantages of gradient-based rank selection on SVD over  1) weight magnitude-based selection without SVD, and 2) brute-force selection of the initial top singular values after SVD (rank-based). Figure \ref{fig:protection} shows that for MRPC and CoLA tasks in the BERT-Base model, the gradient-based approach consistently outperforms both methods. This is because the other two methods lack a direct connection to the loss while the gradient-based approach leverages sensitivity to the training loss, ensuring  better selection of critical parameters, ultimately leading to higher accuracy and robustness.

\subsection{Comparsion with Modern Accelerators}
We evaluate HyFlexPIM against ASADI~\cite{li2024asadi}, ASADI$^\dagger$, SPRINT~\cite{yazdanbakhsh2022sparse}, NMP baseline~\cite{zhou2022transpim}, and a non-PIM baseline using GLUE and WikiText-2 datasets on BERT-Large and GPT-2 models. 
All comparison baselines feature distinct memory hierarchies, compute logic, on-/off-chip memory storage capacity, and in-memory computing capabilities with different benchmarks.
For the fair comparison with ASADI, we maintained the same digital and analog PIM memory capacity as HyFlexPIM; however, ASADI utilizes only SLC, unlike HyFlexPIM, while benefiting from token pruning. The token pruning ratio from \cite{li2024asadi} is applied to reflect its throughput and energy benefits.
SPRINT \cite{yazdanbakhsh2022sparse} reports absolute energy and delay values for each operation including the digital processor, memory access, and in-memory token pruning ratio. These values are utilized to estimate the energy and delay overheads. Non-PIM baseline also exploits the energy and delay numbers of digital processor reported in \cite{yazdanbakhsh2022sparse}, while SRAM and DRAM energy and delay numbers are from simulators described in Section~\ref{baseline}.
The NMP baseline \cite{zhou2022transpim} reports absolute energy values for near-bank processing and HBM access, which are exploited for energy estimation.
Figure \ref{fig:energy} shows linear layer energy efficiency, and Figure \ref{fig:energy_breakdown} (a, c) presents end-to-end energy comparison, while Figure \ref{fig:energy_breakdown} (b, d) provides HyFlexPIM's energy breakdown across different sequence lengths. Additionally, Figure~\ref{fig:speed} compares throughput with ASADI$^\dagger$ and SPRINT. 

\subsubsection{\textbf{HyFlexPIM VS ASADI}}
ASADI introduces RRAM PIM hardware to accelerate end-to-end Transformer operations with FP32, employing a compression method designed to exploit data locality in attention computations. Their focus is primarily on accelerating the attention mechanism, while computing linear layers using analog RRAM PIM exclusively with SLC, without exploiting MLC's superior efficiency. 
On the other hand, HyFlexPIM specifically targets the acceleration of linear layers, which represent a substantial computational bottleneck. 
For linear layer operations (Figure \ref{fig:energy}), our approach achieves maximum $1.24\times$ higher energy efficiency compared to ASADI$^\dagger$ (modified version to use INT8 for linear layers for fair comparison) with 5\% SLC rate, despite ASADI$^\dagger$'s potential accuracy degradation from INT8 quantization. These benefits are more pronounced with shorter sequence lengths where FFN layer becomes the main bottleneck. Considering end-to-end performance, HyFlexPIM achieves 1.1 - 1.86$\times$ speedup compared to ASADI$^\dagger$ (Figure \ref{fig:speed}), primarily due to our 2-bit MLC implementation doubling the throughput with the same energy consumption. At N=1024, we demonstrate maximum $1.67\times$ and $1.44\times$ end-to-end energy efficiency improvements for BERT-Large and GPT-2, respectively (Figure \ref{fig:energy_breakdown} (a,c)), mainly due to ASADI's FP32 precision overhead in attention computations. Furthermore, our approach offers nearly 1.9$\times$ higher area efficiency compared to ASADI$^\dagger$. With approximately 20\% SLC usage, the accuracy drop is less than 1\% in most cases, while achieving an average of 1.23$\times$ improvement in linear layer energy efficiency and 1.79$\times$ speedup.
Notably, HyFlexPIM achieves greater benefits with moderate sequence lengths, aligning well with recent trends in maintaining moderate effective sequence lengths through techniques like RAG models, state-space-based approaches, and selective attention \cite{longmem, beltagy2020longformer, gu2023mamba, infini-attention, retro}.

\subsubsection{\textbf{HyFlexPIM VS SPRINT}}

We compare HyFlexPIM with the SPRINT, which performs in-memory token pruning using RRAM PIM and executes remaining computations using a traditional digital processor. For linear layer operations, as shown in Figure~\ref{fig:energy}, HyFlexPIM achieves maximum 5.4$\times$ energy reduction compared to SPRINT. In terms of end-to-end performance, HyFlexPIM demonstrates maximum 4.94$\times$ and 4.69$\times$ energy efficiency improvements for GLUE and WikiText-2 respectively at N=128 (Figure~\ref{fig:energy_breakdown}), while achieving 10.6$\times$ and 46$\times$ speedup (Figure~\ref{fig:speed}), even when applying SPRINT's 74.6\% sparsity to only SPRINT's attention operations. On average with 20\% SLC rate, HyFlexPIM achieves a 5.34$\times$ reduction in linear layer energy consumption, while providing 10.1$\times$ and 44$\times$ end-to-end speedup for GLUE and WikiText-2 datasets, respectively. The benefits arise from the highly parallel HyFlexPIM architecture and minimized data movements. Note that SPRINT can only reduce the data movement of attention computation. Therefore, the proposed HyFlexPIM obtains higher speedup especially at the lower length of sequences, where the data movement and computation of FFNs dominate. 



\subsubsection{\textbf{HyFlexPIM VS NMP Baseline}}
While NMP reduces data movement cost compared to conventional DRAM-to-cache transfers, it still relies on auxiliary computing units located near the memory, rather than performing computation directly within the memory arrays. As a result, NMP suffers from higher energy and delay overheads from the data movement and digital computation compared to HyFlexPIM.
As shown in Figure \ref{fig:energy}, HyFlexPIM achieves up to 3.31$\times$ higher energy efficiency than the NMP baseline ~\cite{kwon202125} for linear operations. The end-to-end energy efficiency comparison in Figure \ref{fig:energy_breakdown}, demonstrates 4.94$\times$ and 4.69$\times$ improvements for GLUE and WikiText-2, respectively.

\subsubsection{\textbf{HyFlexPIM VS Non-PIM Baseline}}

For linear layer operations, HyFlexPIM achieves a maximum 6.6$\times$ energy efficiency improvement over the non-PIM baseline (Figure \ref{fig:energy}). The end-to-end energy efficiency comparison shows 6.15$\times$ and 5.82$\times$ improvements for GLUE and WikiText-2 respectively (Figure \ref{fig:energy_breakdown} (a, c)). These substantial gains are attributed to two main factors in the baseline: high energy costs from frequent DRAM off-chip data movements and inefficient GEMV operations in conventional digital processors without the benefits of PIM.

\begin{figure}[th]
\centering
\includegraphics[width=1\columnwidth]{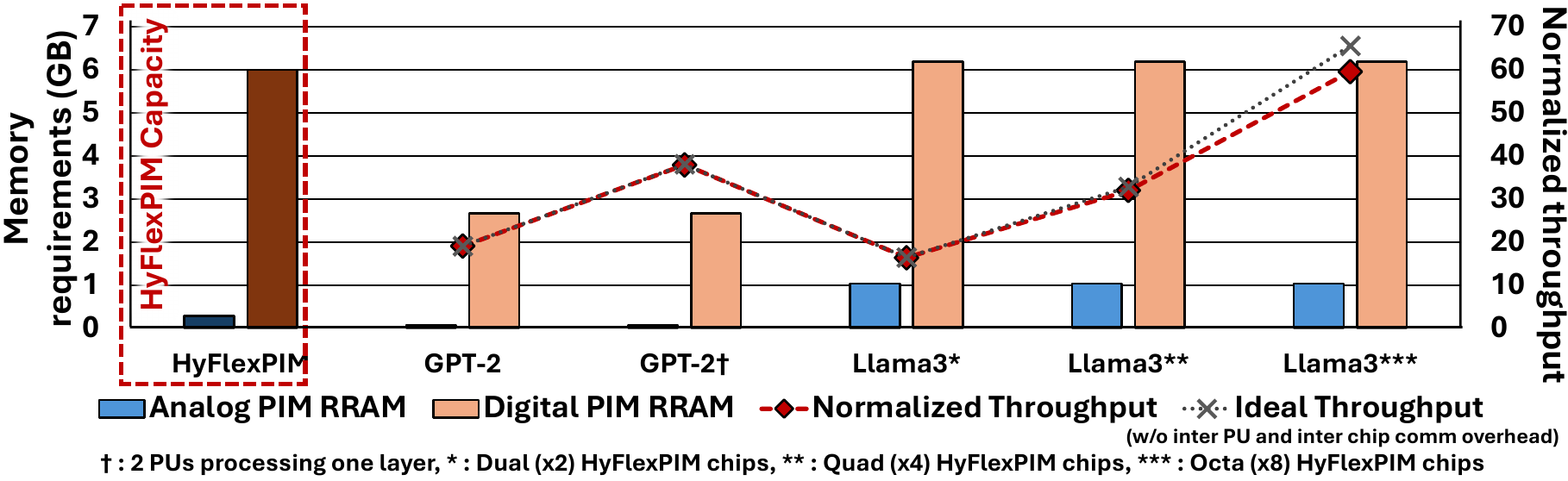} 
\caption{Memory requirements and throughput scalability of HyFlexPIM for N = 8192.}
\label{fig:ratio}
\end{figure}

\subsubsection{\textbf{HyFlexPIM scalability}}

As described in Section~\ref{sec: hardware_arch}, HyFlexPIM provides flexibility and scalability through 1) tensor parallelism, where multiple PUs collaborate to enhance throughput, and 2) pipeline parallelism across multiple chips. For example, tensor parallelism is applied to smaller models such as GPT-2, where two PUs collaboratively process a single layer, improving throughput by 1.99$\times$ compared to the case without parallelism, as shown in Figure~\ref{fig:ratio}. This results in only a minor degradation compared to the ideal 2$\times$ scalability, attributed to the PU-to-PU partial sum transfer overhead. Conversely, for large-scale models such as Llama3, a single PU cannot hold each layer due to its large hidden dimension so that two PUs process a layer. It also requires pipeline parallelism as a chip cannot accommodate the entire model, necessitating two chips at minimum. We analyze Llama3 dual/quad/octa-chip configurations. Quad and octa chips achieve 1.96$\times$ and 3.65$\times$ throughput improvements compared to the dual-core case including the chip-to-chip communication overhead to deliver the hidden layer to the next chip. This results in only a minor degradation compared to the ideal 2$\times$ and 4$\times$ scaling benefits. 

\section{Related Works}
Over the last decade, RRAM PIM has been explored to the neural network acceleration~\cite{rram_isaac, huang2023nonvolatile, rram_prime, yu2021rram, li2024asadi, yazdanbakhsh2022sparse, liu2023hardsea}. The pioneering works, ISAAC~\cite{rram_isaac} and \cite{huang2023nonvolatile} demonstrate a pipelined RRAM crossbar and hybrid RRAM capable of efficiently executing CNNs. PRIME~\cite{rram_prime} introduces a reconfigurable RRAM PIM architecture, where the same PIM hardware can be configured as both regular memory and computing elements. More recently, RRAM PIM has been applied to accelerate Transformer models. For instance, SPRINT employs analog RRAM PIM to compute correlation scores as a pre-processor for pruning unimportant tokens in the attention mechanism~\cite{yazdanbakhsh2022sparse}. However, the primary computation is still carried out by energy-intensive digital processors. Moreover, SPRINT only accelerates the attention mechanism, leaving other computation- and memory-intensive tasks, such as FFNs, unoptimized. Similarly, HARDSEA~\cite{liu2023hardsea} utilizes analog RRAM PIM solely to predict token relevance. ASADI~\cite{li2024asadi} introduces a hybrid analog-digital RRAM PIM hardware and a compression method designed to exploit data locality in attention computing. However, ASADI relies exclusively on SLC RRAM, thereby missing the potential benefits of a high-density and energy-efficient MLC RRAM PIM architecture. Importantly, most existing RRAM PIM approaches depend on the inherent error resilience of neural networks to circumvent the challenges posed by noisy analog RRAM PIM operations. In contrast, our approach seeks to proactively reshape the model to enhance its robustness, with principles that can be broadly applied to other memory topologies.


\section{Conclusion}

Analog processing-in-memory (PIM) offers significant potential for energy and speed improvements in Transformer acceleration through parallel processing and multi-level cell (MLC) capabilities, while digital PIM provides reliable computation for dynamic weights updated in real-time. Despite these advantages, deploying PIM effectively for sophisticated models such as Transformers remains challenging due to various practical considerations. To maximize the benefits of both domains, we propose a mixed-signal approach that strategically assigns operations based on their characteristics, utilizing digital PIM for dynamic weights and analog PIM with both SLC and MLC for static weight processing.

Rather than relying passively on inherent error resiliency, our gradient redistribution technique proactively reshapes models to be more compatible with the hybrid hardware platform. This approach enables us to maximize MLC utilization over SLC in analog PIM while preserving critical computations in SLC, achieving efficient processing while maintaining model accuracy. We expect this comprehensive approach will enable practical deployment of mixed-signal PIM technology in future large-scale AI systems.




\section*{Acknowledgements}
This work was supported in part by PRISM and CoCoSys, centers in
JUMP 2.0, an SRC program sponsored by DARPA \#434690, and by the National
Research Foundation of Korea (MSIT) grants RS-2024-00347090 and
RS-2024-00405857.

\bibliographystyle{plain}
\bibliography{refs.bib}

\begin{thebibliography}{10}

\bibitem{abdi2007singular}
Herv{\'e} Abdi.
\newblock Singular value decomposition (svd) and generalized singular value decomposition.
\newblock {\em Encyclopedia of measurement and statistics}, 907(912):44, 2007.

\bibitem{agrawal2019xcel}
Amogh Agrawal, Akhilesh Jaiswal, Deboleena Roy, Bing Han, Gopalakrishnan Srinivasan, Aayush Ankit, and Kaushik Roy.
\newblock Xcel-ram: Accelerating binary neural networks in high-throughput sram compute arrays.
\newblock {\em IEEE Transactions on Circuits and Systems I: Regular Papers}, 66(8):3064--3076, 2019.

\bibitem{llama3.2}
Meta AI.
\newblock Llama 3.2: Multilingual large language models.
\newblock \url{https://www.llama.com}, 2024.

\bibitem{ali2019memory}
Mustafa~F Ali, Akhilesh Jaiswal, and Kaushik Roy.
\newblock In-memory low-cost bit-serial addition using commodity dram technology.
\newblock {\em IEEE Transactions on Circuits and Systems I: Regular Papers}, 67(1):155--165, 2019.

\bibitem{andrulis2023raella}
Tanner Andrulis, Joel~S Emer, and Vivienne Sze.
\newblock Raella: Reforming the arithmetic for efficient, low-resolution, and low-loss analog pim: No retraining required!
\newblock In {\em Proceedings of the 50th Annual International Symposium on Computer Architecture}, pages 1--16, 2023.

\bibitem{mem_compiler}
ARM.
\newblock {Artisan Memory Compilers}.
\newblock \url{https://developer.arm.com/ip-products/physical-ip/embedded-memory}, 2021.
\newblock Accessed: 2021-11-08.

\bibitem{beltagy2020longformer}
Iz~Beltagy, Matthew~E Peters, and Arman Cohan.
\newblock Longformer: The long-document transformer.
\newblock {\em arXiv preprint arXiv:2004.05150}, 2020.

\bibitem{retro}
Sebastian Borgeaud, Arthur Mensch, Jordan Hoffmann, Trevor Cai, Eliza Rutherford, Katie Millican, George~Bm Van Den~Driessche, Jean-Baptiste Lespiau, Bogdan Damoc, Aidan Clark, et~al.
\newblock Improving language models by retrieving from trillions of tokens.
\newblock In {\em International conference on machine learning}, pages 2206--2240. PMLR, 2022.

\bibitem{cao2021neural}
Weidong Cao, Yilong Zhao, Adith Boloor, Yinhe Han, Xuan Zhang, and Li~Jiang.
\newblock Neural-pim: Efficient processing-in-memory with neural approximation of peripherals.
\newblock {\em IEEE Transactions on Computers}, 71(9):2142--2155, 2021.

\bibitem{10447756}
Zhiyang Chen, Yousong Zhu, Zhaowen Li, Fan Yang, Chaoyang Zhao, Jinqiao Wang, and Ming Tang.
\newblock The devil is in details: Delving into lite ffn design for vision transformers.
\newblock In {\em ICASSP 2024 - 2024 IEEE International Conference on Acoustics, Speech and Signal Processing (ICASSP)}, pages 4130--4134, 2024.

\bibitem{rram_prime}
Ping Chi, Shuangchen Li, Cong Xu, Tao Zhang, Jishen Zhao, Yongpan Liu, Yu~Wang, and Yuan Xie.
\newblock Prime: a novel processing-in-memory architecture for neural network computation in reram-based main memory.
\newblock In {\em Proceedings of the 43rd International Symposium on Computer Architecture}, ISCA '16, pages 27–--39. IEEE Press, 2016.

\bibitem{devlin2018bert}
Jacob Devlin.
\newblock Bert: Pre-training of deep bidirectional transformers for language understanding.
\newblock {\em arXiv preprint arXiv:1810.04805}, 2018.

\bibitem{6218223}
Xiangyu Dong, Cong Xu, Yuan Xie, and Norman~P. Jouppi.
\newblock Nvsim: A circuit-level performance, energy, and area model for emerging nonvolatile memory.
\newblock {\em IEEE Transactions on Computer-Aided Design of Integrated Circuits and Systems}, 31(7):994--1007, 2012.

\bibitem{dosovitskiy2021image}
Alexey Dosovitskiy, Lucas Beyer, Alexander Kolesnikov, Dirk Weissenborn, Xiaohua Zhai, Thomas Unterthiner, Mostafa Dehghani, Matthias Minderer, Georg Heigold, Sylvain Gelly, Jakob Uszkoreit, and Neil Houlsby.
\newblock An image is worth 16x16 words: Transformers for image recognition at scale.
\newblock In {\em International Conference on Learning Representations}, 2021.

\bibitem{fan2024efficientopenmodificationspectral}
Keming Fan, Wei-Chen Chen, Sumukh Pinge, H.~S.~Philip Wong, and Tajana Rosing.
\newblock Efficient open modification spectral library searching in high-dimensional space with multi-level-cell memory, 2024.

\bibitem{gao2019computedram}
Fei Gao, Georgios Tziantzioulis, and David Wentzlaff.
\newblock Computedram: In-memory compute using off-the-shelf drams.
\newblock In {\em Proceedings of the 52nd annual IEEE/ACM international symposium on microarchitecture}, page 100–113, 2019.

\bibitem{JSSC_SGD}
Sujan~K Gonugondla, Mingu Kang, and Naresh~R Shanbhag.
\newblock A variation-tolerant in-memory machine learning classifier via on-chip training.
\newblock {\em JSSC}, 53(11):3163--3173, November 2018.

\bibitem{8640245}
Alessandro Grossi, Elisa Vianello, Mohamed~M. Sabry, Marios Barlas, Laurent Grenouillet, Jean Coignus, Edith Beigne, Tony Wu, Binh~Q. Le, Mary~K. Wootters, Cristian Zambelli, Etienne Nowak, and Subhasish Mitra.
\newblock Resistive ram endurance: Array-level characterization and correction techniques targeting deep learning applications.
\newblock {\em IEEE Transactions on Electron Devices}, 66(3):1281--1288, 2019.

\bibitem{gu2023mamba}
Albert Gu and Tri Dao.
\newblock Mamba: Linear-time sequence modeling with selective state spaces.
\newblock {\em arXiv preprint arXiv:2312.00752}, 2023.

\bibitem{huang2023nonvolatile}
Wei-Hsing Huang, Tai-Hao Wen, Je-Min Hung, Win-San Khwa, Yun-Chen Lo, Chuan-Jia Jhang, Huna-Hsi Hsu, Yu-Hsiana Chin, Yu-Chiao Chen, Chuna-Chuan Lo, et~al.
\newblock A nonvolatile al-edge processor with 4mb slc-mlc hybrid-mode reram compute-in-memory macro and 51.4-251tops/w.
\newblock In {\em 2023 IEEE International Solid-State Circuits Conference (ISSCC)}, pages 15--17. IEEE, 2023.

\bibitem{hung2021four}
Je-Min Hung, Cheng-Xin Xue, Hui-Yao Kao, Yen-Hsiang Huang, Fu-Chun Chang, Sheng-Po Huang, Ta-Wei Liu, Chuan-Jia Jhang, Chin-I Su, Win-San Khwa, et~al.
\newblock A four-megabit compute-in-memory macro with eight-bit precision based on cmos and resistive random-access memory for ai edge devices.
\newblock {\em Nature Electronics}, 4(12):921--930, 2021.

\bibitem{imani2019floatpim}
Mohsen Imani, Saransh Gupta, Yeseong Kim, and Tajana Rosing.
\newblock Floatpim: In-memory acceleration of deep neural network training with high precision.
\newblock In {\em Proceedings of the 46th International Symposium on Computer Architecture}, pages 802--815, 2019.

\bibitem{jiang2019circuit}
Yuning Jiang, Peng Huang, Zheng Zhou, and Jinfeng Kang.
\newblock Circuit design of rram-based neuromorphic hardware systems for classification and modified hebbian learning.
\newblock {\em Science China Information Sciences}, 62:1--19, 2019.

\bibitem{jin2021rehy}
Hai Jin, Cong Liu, Haikun Liu, Ruikun Luo, Jiahong Xu, Fubing Mao, and Xiaofei Liao.
\newblock Rehy: A reram-based digital/analog hybrid pim architecture for accelerating cnn training.
\newblock {\em IEEE Transactions on Parallel and Distributed Systems}, 33(11):2872–2884, 2021.

\bibitem{9663036}
Hai Jin, Cong Liu, Haikun Liu, Ruikun Luo, Jiahong Xu, Fubing Mao, and Xiaofei Liao.
\newblock Rehy: A reram-based digital/analog hybrid pim architecture for accelerating cnn training.
\newblock {\em IEEE Transactions on Parallel and Distributed Systems}, 33(11):2872--2884, 2022.

\bibitem{jouppi2021ten}
Norman~P Jouppi, Doe~Hyun Yoon, Matthew Ashcraft, Mark Gottscho, Thomas~B Jablin, George Kurian, James Laudon, Sheng Li, Peter Ma, Xiaoyu Ma, et~al.
\newblock Ten lessons from three generations shaped google’s tpuv4i: Industrial product.
\newblock In {\em 2021 ACM/IEEE 48th Annual International Symposium on Computer Architecture (ISCA)}, pages 1--14. IEEE, 2021.

\bibitem{kim2023samba}
Dong~Eun Kim, Aayush Ankit, Cheng Wang, and Kaushik Roy.
\newblock Samba: sparsity aware in-memory computing based machine learning accelerator.
\newblock {\em IEEE Transactions on Computers}, 72(9):2615--2627, 2023.

\bibitem{Krizhevsky09learningmultiple}
Alex Krizhevsky.
\newblock Learning multiple layers of features from tiny images.
\newblock Technical report, 2009.

\bibitem{kull20133}
Lukas Kull, Thomas Toifl, Martin Schmatz, Pier~Andrea Francese, Christian Menolfi, Matthias Braendli, Marcel Kossel, Thomas Morf, Toke~Meyer Andersen, and Yusuf Leblebici.
\newblock A 3.1 mw 8b 1.2 gs/s single-channel asynchronous sar adc with alternate comparators for enhanced speed in 32 nm digital soi cmos.
\newblock {\em IEEE Journal of Solid-State Circuits}, 48(12):3049--3058, 2013.

\bibitem{kwon202125}
Young-Cheon Kwon, Suk~Han Lee, Jaehoon Lee, Sang-Hyuk Kwon, Je~Min Ryu, Jong-Pil Son, O~Seongil, Hak-Soo Yu, Haesuk Lee, Soo~Young Kim, et~al.
\newblock 25.4 a 20nm 6gb function-in-memory dram, based on hbm2 with a 1.2 tflops programmable computing unit using bank-level parallelism, for machine learning applications.
\newblock In {\em 2021 IEEE International Solid-State Circuits Conference (ISSCC)}, volume~64, pages 350--352. IEEE, 2021.

\bibitem{li2024asadi}
Huize Li, Zhaoying Li, Zhenyu Bai, and Tulika Mitra.
\newblock Asadi: Accelerating sparse attention using diagonal-based in-situ computing.
\newblock In {\em 2024 IEEE International Symposium on High-Performance Computer Architecture (HPCA)}, pages 774--787. IEEE, 2024.

\bibitem{liu2023hardsea}
Shiwei Liu, Chen Mu, Hao Jiang, Yunzhengmao Wang, Jinshan Zhang, Feng Lin, Keji Zhou, Qi~Liu, and Chixiao Chen.
\newblock Hardsea: Hybrid analog-reram clustering and digital-sram in-memory computing accelerator for dynamic sparse self-attention in transformer.
\newblock {\em IEEE Transactions on Very Large Scale Integration (VLSI) Systems}, 2023.

\bibitem{liu2023area}
Shuang Liu, JJ~Wang, JT~Zhou, SG~Hu, Qi~Yu, TP~Chen, and Yang Liu.
\newblock An area-and energy-efficient spiking neural network with spike-time-dependent plasticity realized with sram processing-in-memory macro and on-chip unsupervised learning.
\newblock {\em IEEE Transactions on Biomedical Circuits and Systems}, 17(1):92--104, 2023.

\bibitem{lu2020hardware}
Siyuan Lu, Meiqi Wang, Shuang Liang, Jun Lin, and Zhongfeng Wang.
\newblock Hardware accelerator for multi-head attention and position-wise feed-forward in the transformer.
\newblock In {\em 2020 IEEE 33rd International System-on-Chip Conference (SOCC)}, pages 84--89. IEEE, 2020.

\bibitem{lv2023lightformer}
Xiuqing Lv, Peng Zhang, Sunzhu Li, Guobing Gan, and Yueheng Sun.
\newblock Lightformer: Light-weight transformer using svd-based weight transfer and parameter sharing.
\newblock In {\em Findings of the Association for Computational Linguistics: ACL 2023}, pages 10323--10335, 2023.

\bibitem{marcus-etal-1993-building}
Mitchell~P. Marcus, Beatrice Santorini, and Mary~Ann Marcinkiewicz.
\newblock Building a large annotated corpus of {E}nglish: The {P}enn {T}reebank.
\newblock {\em Computational Linguistics}, 19(2):313--330, 1993.

\bibitem{wikitext2}
Stephen Merity.
\newblock {The WikiText Long Term Dependency Language Modeling Dataset}.
\newblock \url{https://blog.salesforceairesearch.com/the-wikitext-long-term-dependency-language-modeling-dataset/}, 2021.
\newblock Accessed: 2021-11-08.

\bibitem{merity2016pointer}
Stephen Merity, Caiming Xiong, James Bradbury, and Richard Socher.
\newblock Pointer sentinel mixture models, 2016.

\bibitem{infini-attention}
Tsendsuren Munkhdalai, Manaal Faruqui, and Siddharth Gopal.
\newblock Leave no context behind: Efficient infinite context transformers with infini-attention.
\newblock {\em arXiv preprint arXiv:2404.07143}, 2024.

\bibitem{adc_survey}
Boris Murmann.
\newblock {ADC Performance Survey 1997-2024}.
\newblock [Online]. Available: \url{https://github.com/bmurmann/ADC-survey}.

\bibitem{o2017fine}
Mike O'Connor, Niladrish Chatterjee, Donghyuk Lee, John Wilson, Aditya Agrawal, Stephen~W Keckler, and William~J Dally.
\newblock Fine-grained dram: Energy-efficient dram for extreme bandwidth systems.
\newblock In {\em Proceedings of the 50th Annual IEEE/ACM International Symposium on Microarchitecture}, pages 41--54, 2017.

\bibitem{prezioso2015training}
Mirko Prezioso, Farnood Merrikh-Bayat, Brian~D Hoskins, Gina~C Adam, Konstantin~K Likharev, and Dmitri~B Strukov.
\newblock Training and operation of an integrated neuromorphic network based on metal-oxide memristors.
\newblock {\em Nature}, 521(7550):61--64, 2015.

\bibitem{10.1145/3579371.3589057}
Yubin Qin, Yang Wang, Dazheng Deng, Zhiren Zhao, Xiaolong Yang, Leibo Liu, Shaojun Wei, Yang Hu, and Shouyi Yin.
\newblock Fact: Ffn-attention co-optimized transformer architecture with eager correlation prediction.
\newblock In {\em Proceedings of the 50th Annual International Symposium on Computer Architecture}, ISCA '23, New York, NY, USA, 2023. Association for Computing Machinery.

\bibitem{Radford2019LanguageMA}
A.~Radford, Jeffrey Wu, R.~Child, David Luan, Dario Amodei, and Ilya Sutskever.
\newblock Language models are unsupervised multitask learners.
\newblock 2019.

\bibitem{radford2019language}
Alec Radford, Jeffrey Wu, Rewon Child, David Luan, Dario Amodei, Ilya Sutskever, et~al.
\newblock Language models are unsupervised multitask learners.
\newblock {\em OpenAI blog}, 1(8):9, 2019.

\bibitem{ramadan2019adaptive}
Misbah Ramadan, Nicol{\'a}s Wainstein, Ran Ginosar, and Shahar Kvatinsky.
\newblock Adaptive programming in multi-level cell reram.
\newblock {\em Microelectronics Journal}, 90:169--180, 2019.

\bibitem{RAMADAN2019169}
Misbah Ramadan, Nicol{\'a}s Wainstein, Ran Ginosar, and Shahar Kvatinsky.
\newblock Adaptive programming in multi-level cell reram.
\newblock {\em Microelectronics Journal}, 90:169--180, 2019.

\bibitem{ren2023associative}
Yiming Ren, Bobo Tian, Mengge Yan, Guangdi Feng, Bin Gao, Fangyu Yue, Hui Peng, Xiaodong Tang, Qiuxiang Zhu, Junhao Chu, et~al.
\newblock Associative learning of a three-terminal memristor network for digits recognition.
\newblock {\em Science China Information Sciences}, 66(2):122403, 2023.

\bibitem{saberi2011analysis}
Mehdi Saberi, Reza Lotfi, Khalil Mafinezhad, and Wouter~A Serdijn.
\newblock Analysis of power consumption and linearity in capacitive digital-to-analog converters used in successive approximation adcs.
\newblock {\em IEEE Transactions on Circuits and Systems I: Regular Papers}, 58(8):1736--1748, 2011.

\bibitem{rram_isaac}
Ali Shafiee, Anirban Nag, Naveen Muralimanohar, Rajeev Balasubramonian, John~Paul Strachan, Miao Hu, R.~Stanley Williams, and Vivek Srikumar.
\newblock Isaac: a convolutional neural network accelerator with in-situ analog arithmetic in crossbars.
\newblock In {\em Proceedings of the 43rd International Symposium on Computer Architecture}, ISCA '16, page 14–26. IEEE Press, 2016.

\bibitem{10.1145/3007787.3001139}
Ali Shafiee, Anirban Nag, Naveen Muralimanohar, Rajeev Balasubramonian, John~Paul Strachan, Miao Hu, R.~Stanley Williams, and Vivek Srikumar.
\newblock Isaac: a convolutional neural network accelerator with in-situ analog arithmetic in crossbars.
\newblock {\em SIGARCH Comput. Archit. News}, 44(3):14–26, jun 2016.

\bibitem{shafiee2016isaac}
Ali Shafiee, Anirban Nag, Naveen Muralimanohar, Rajeev Balasubramonian, John~Paul Strachan, Miao Hu, R~Stanley Williams, and Vivek Srikumar.
\newblock Isaac: A convolutional neural network accelerator with in-situ analog arithmetic in crossbars.
\newblock {\em ACM SIGARCH Computer Architecture News}, 44(3):14--26, 2016.

\bibitem{sharma2020pci}
Debendra~Das Sharma.
\newblock Pci express{\textregistered} 6.0 specification at 64.0 gt/s with pam-4 signaling: a low latency, high bandwidth, high reliability and cost-effective interconnect.
\newblock In {\em 2020 IEEE Symposium on High-Performance Interconnects (HOTI)}, pages 1--8. IEEE, 2020.

\bibitem{smullen2011relaxing}
Clinton~W Smullen, Vidyabhushan Mohan, Anurag Nigam, Sudhanva Gurumurthi, and Mircea~R Stan.
\newblock Relaxing non-volatility for fast and energy-efficient stt-ram caches.
\newblock In {\em 2011 IEEE 17th International Symposium on High Performance Computer Architecture}, pages 50--61. IEEE, 2011.

\bibitem{song202452}
Chang~Eun Song, Yidong Li, Amardeep Ramnani, Pulkit Agrawal, Purvi Agrawal, Sung-Joon Jang, Sang-Seol Lee, Tajana Rosing, and Mingu Kang.
\newblock 52.5 tops/w 1.7 ghz reconfigurable xgboost inference accelerator based on modular-unit-tree with dynamic data and compute gating.
\newblock In {\em 2024 IEEE Custom Integrated Circuits Conference (CICC)}, pages 1--2. IEEE, 2024.

\bibitem{10.1145/3665314.3670798}
Chang~Eun Song, Ashkan Moradifirouzabadi, Tajana Rosing, and Mingu Kang.
\newblock Efficient transformer acceleration via reconfiguration for encoder and decoder models and sparsity-aware algorithm mapping.
\newblock In {\em Proceedings of the 29th ACM/IEEE International Symposium on Low Power Electronics and Design}, ISLPED '24, page 1–6, New York, NY, USA, 2024. Association for Computing Machinery.

\bibitem{stillmaker2017scaling}
Aaron Stillmaker and Bevan Baas.
\newblock Scaling equations for the accurate prediction of cmos device performance from 180 nm to 7 nm.
\newblock {\em Integration}, 58:74--81, 2017.

\bibitem{talati2016logic}
Nishil Talati, Saransh Gupta, Pravin Mane, and Shahar Kvatinsky.
\newblock Logic design within memristive memories using memristor-aided logic (magic).
\newblock {\em IEEE Transactions on Nanotechnology}, 15(4):635–650, 2016.

\bibitem{topal2021exploring}
M~Onat Topal, Anil Bas, and Imke van Heerden.
\newblock Exploring transformers in natural language generation: Gpt, bert, and xlnet.
\newblock {\em arXiv preprint arXiv:2102.08036}, 2021.

\bibitem{truong2021racer}
Minh~SQ Truong, Eric Chen, Deanyone Su, Liting Shen, Alexander Glass, L~Richard Carley, James~A Bain, and Saugata Ghose.
\newblock Racer: Bit-pipelined processing using resistive memory.
\newblock In {\em MICRO-54: 54th Annual IEEE/ACM International Symposium on Microarchitecture}, page 100–116, 2021.

\bibitem{wan202033}
Weier Wan, Rajkumar Kubendran, S~Burc Eryilmaz, Wenqiang Zhang, Yan Liao, Dabin Wu, Stephen Deiss, Bin Gao, Priyanka Raina, Siddharth Joshi, et~al.
\newblock 33.1 a 74 tmacs/w cmos-rram neurosynaptic core with dynamically reconfigurable dataflow and in-situ transposable weights for probabilistic graphical models.
\newblock In {\em 2020 IEEE International Solid-State Circuits Conference-(ISSCC)}, pages 498--500. IEEE, 2020.

\bibitem{wang2018glue}
Alex Wang.
\newblock Glue: A multi-task benchmark and analysis platform for natural language understanding.
\newblock {\em arXiv preprint arXiv:1804.07461}, 2018.

\bibitem{wang2019glue}
Alex Wang, Amanpreet Singh, Julian Michael, Felix Hill, Omer Levy, and Samuel~R. Bowman.
\newblock Glue: A multi-task benchmark and analysis platform for natural language understanding, 2019.

\bibitem{wang2020hat}
Hanrui Wang, Zhanghao Wu, Zhijian Liu, Han Cai, Ligeng Zhu, Chuang Gan, and Song Han.
\newblock Hat: Hardware-aware transformers for efficient natural language processing.
\newblock {\em arXiv preprint arXiv:2005.14187}, 2020.

\bibitem{longmem}
Weizhi Wang, Li~Dong, Hao Cheng, Xiaodong Liu, Xifeng Yan, Jianfeng Gao, and Furu Wei.
\newblock Augmenting language models with long-term memory.
\newblock {\em Advances in Neural Information Processing Systems}, 36, 2024.

\bibitem{wang2024beacongnn}
Yuyue Wang, Xiurui Pan, Yuda An, Jie Zhang, and Glenn Reinman.
\newblock Beacongnn: Large-scale gnn acceleration with out-of-order streaming in-storage computing.
\newblock In {\em 2024 IEEE International Symposium on High-Performance Computer Architecture (HPCA)}, pages 330--344. IEEE, 2024.

\bibitem{wong2010phase}
H-S~Philip Wong, Simone Raoux, SangBum Kim, Jiale Liang, John~P Reifenberg, Bipin Rajendran, Mehdi Asheghi, and Kenneth~E Goodson.
\newblock Phase change memory.
\newblock {\em Proceedings of the IEEE}, 98(12):2201--2227, 2010.

\bibitem{wu2024pim}
Yuting Wu, Ziyu Wang, and Wei~D Lu.
\newblock Pim gpt a hybrid process in memory accelerator for autoregressive transformers.
\newblock {\em npj Unconventional Computing}, 1(1):4, 2024.

\bibitem{xue202116}
Cheng-Xin Xue, Je-Min Hung, Hui-Yao Kao, Yen-Hsiang Huang, Sheng-Po Huang, Fu-Chun Chang, Peng Chen, Ta-Wei Liu, Chuan-Jia Jhang, Chin-I Su, et~al.
\newblock 16.1 a 22nm 4mb 8b-precision reram computing-in-memory macro with 11.91 to 195.7 tops/w for tiny ai edge devices.
\newblock In {\em 2021 IEEE International Solid-State Circuits Conference (ISSCC)}, volume~64, pages 245--247. IEEE, 2021.

\bibitem{yang2023processing}
Guowei Yang, Cansu Demirkiran, Zeynep~Ece Kizilates, Carlos A~R{\'\i}os Ocampo, Ayse~K Coskun, and Ajay Joshi.
\newblock Processing-in-memory using optically-addressed phase change memory.
\newblock In {\em 2023 IEEE/ACM International Symposium on Low Power Electronics and Design (ISLPED)}, pages 1--6. IEEE, 2023.

\bibitem{yang2024fsl}
Haichao Yang, Chang~Eun Song, Weihong Xu, Behnam Khaleghi, Uday Mallappa, Monil Shah, Keming Fan, Mingu Kang, and Tajana Rosing.
\newblock Fsl-hdnn: A 5.7 tops/w end-to-end few-shot learning classifier accelerator with feature extraction and hyperdimensional computing.
\newblock In {\em 2024 IEEE European Solid-State Electronics Research Conference (ESSERC)}, pages 33--36. IEEE, 2024.

\bibitem{yang2020retransformer}
Xiaoxuan Yang, Bonan Yan, Hai Li, and Yiran Chen.
\newblock Retransformer: Reram-based processing-in-memory architecture for transformer acceleration.
\newblock In {\em Proceedings of the 39th International Conference on Computer-Aided Design}, pages 1--9, 2020.

\bibitem{10.1145/3400302.3415640}
Xiaoxuan Yang, Bonan Yan, Hai Li, and Yiran Chen.
\newblock Retransformer: Reram-based processing-in-memory architecture for transformer acceleration.
\newblock In {\em Proceedings of the 39th International Conference on Computer-Aided Design}, ICCAD '20, New York, NY, USA, 2020. Association for Computing Machinery.

\bibitem{yao2020fully}
Peng Yao, Huaqiang Wu, Bin Gao, Jianshi Tang, Qingtian Zhang, Wenqiang Zhang, J~Joshua Yang, and He~Qian.
\newblock Fully hardware-implemented memristor convolutional neural network.
\newblock {\em Nature}, 577(7792):641--646, 2020.

\bibitem{yazdanbakhsh2022sparse}
Amir Yazdanbakhsh, Ashkan Moradifirouzabadi, Zheng Li, and Mingu Kang.
\newblock Sparse attention acceleration with synergistic in-memory pruning and on-chip recomputation.
\newblock In {\em 2022 55th IEEE/ACM International Symposium on Microarchitecture (MICRO)}, pages 744--762. IEEE, 2022.

\bibitem{yoshioka2024visiontransformers}
Kentaro Yoshioka.
\newblock vision-transformers-cifar10: Training vision transformers (vit) and related models on cifar-10.
\newblock \url{https://github.com/kentaroy47/vision-transformers-cifar10}, 2024.

\bibitem{yu2021rram}
Shimeng Yu, Wonbo Shim, Xiaochen Peng, and Yandong Luo.
\newblock Rram for compute-in-memory: From inference to training.
\newblock {\em IEEE Transactions on Circuits and Systems I: Regular Papers}, 68(7):2753--2765, 2021.

\bibitem{zhang2020neuro}
Wenqiang Zhang, Bin Gao, Jianshi Tang, Peng Yao, Shimeng Yu, Meng-Fan Chang, Hoi-Jun Yoo, He~Qian, and Huaqiang Wu.
\newblock Neuro-inspired computing chips.
\newblock {\em Nature electronics}, 3(7):371--382, 2020.

\bibitem{zhou2022transpim}
Minxuan Zhou, Weihong Xu, Jaeyoung Kang, and Tajana Rosing.
\newblock Transpim: A memory-based acceleration via software-hardware co-design for transformer.
\newblock In {\em 2022 IEEE International Symposium on High-Performance Computer Architecture (HPCA)}, pages 1071--1085. IEEE, 2022.

\bibitem{zidan2018future}
Mohammed~A Zidan, John~Paul Strachan, and Wei~D Lu.
\newblock The future of electronics based on memristive systems.
\newblock {\em Nature electronics}, 1(1):22--29, 2018.

\end{thebibliography}

\end{document}